\documentclass[a4paper,11pt]{article}
\pdfoutput=1
\usepackage{jheppub}
\usepackage[T1]{fontenc}
\usepackage{diagbox}
\usepackage{adjustbox}
\usepackage{multirow}
\usepackage{pifont}
\usepackage{array}
\usepackage{mathtools}
\usepackage{booktabs}
\usepackage[dvipsnames]{xcolor}
\usepackage{tcolorbox}
\usepackage{float}
\usepackage{placeins}
\newcommand{\eqal}[1]{\begin{align}#1\end{align}}

\usepackage{verbatim}

\newcommand{\yong}[1]{{\color{black}#1}}

\begin{document}
\title{\boldmath \yong{Electroweak} Phase Transition and Gravitational Waves in the Type-II Seesaw Model}

\author[a]{Ruiyu Zhou,}
\emailAdd{zhoury@cqupt.edu.cn}
\author[b,c]{Ligong Bian\footnote{Corresponding author.},}
\emailAdd{lgbycl@cqu.edu.cn}
\author[d]{Yong Du}
\emailAdd{yongdu@itp.ac.cn}

\affiliation[a]{School of Science, Chongqing University of Posts and Telecommunications, Chongqing 400065, P. R. China}
\affiliation[b]{Department of Physics and Chongqing Key Laboratory for Strongly Coupled Physics, Chongqing University, Chongqing 401331, P. R. China}

\affiliation[c]{
Center for High Energy Physics, Peking University, Beijing 100871, P. R. China}

\affiliation[d]{CAS Key Laboratory of Theoretical Physics, Institute of Theoretical Physics, Chinese Academy of Sciences, Beijing 100190, P. R. China}

\abstract{The type-II seesaw model is a possible candidate for simultaneously explaining non-vanishing neutrino masses and the observed baryon asymmetry of the Universe. In this work, we study in detail the pattern of phase transition and the gravitational wave production of this model. We find a strong first-order \yong{electroweak} phase transition generically prefers positive Higgs portal couplings and a light triplet below $\sim550$\,GeV. In addition, we find the gravitational wave yield generated during the phase transition \yong{would be} at the edge of BBO sensitivity and could be further examined by Ultimate-DECIGO.}

\maketitle

\section{Introduction}

\yong{The discovery of the Higgs boson in 2012\,\cite{ATLAS:2012yve,CMS:2012qbp} completes the picture of the Standard Model (SM). However, within the SM framework, it is recognized that neutrinos are exactly massless particles as a result of a global U(1)$_\ell$ symmetry, conflicting with the observed phenomena of neutrino oscillations\,\cite{Fukuda:1998mi,Ahmad:2001an}. Furthermore, the phase transition with the observed 125\,GeV Higgs boson in the SM will be of a crossover type\,\cite{Kajantie:1995kf,Kajantie:1996mn,Kajantie:1996qd}, thus making the SM inadequate to explain the observed asymmetry of baryons\,\cite{Planck:2018vyg}. Both these facts, together with some other fundamental questions like the nature of dark matter, imply that the SM cannot be the complete theory and extension of it is needed.}

\yong{Among those extensions of the SM that can be responsible for massive neutrinos, the type-I, -II and -III seesaw models\,\cite{Minkowski:1977sc,Ramond:1979py,GellMann:1980vs,Yanagida:1979as,Mohapatra:1979ia,Schechter:1980gr,Schechter:1981cv,Konetschny:1977bn,Cheng:1980qt,Lazarides:1980nt,Magg:1980ut,Foot:1988aq,Witten:1985bz,Mohapatra:1986aw,Mohapatra:1986bd,Val86,Barr:2003nn,Mohapatra:1980yp}, inspired by the pioneering work of Weinberg\,\cite{Weinberg:1979sa}, have been extensively studied as they can naturally induce neutrino masses through the seesaw mechanism. In particular, all these three models predict the violation of lepton numbers by two units in contrast to their conservation in the SM. While it is not yet clear which mechanism is realized in practice, nowadays it is widely known that the type-I and -III models would be well beyond the reach of current experiments due to the largeness of the seesaw scales. In contrast, allowing the neutrino Yukawa couplings to be tiny, low-scale type-I and -III seesaw models would become possible and have also been investigated in literatures\,\cite{Han:2006ip,Atre:2009rg,FileviezPerez:2009hdc,Alva:2014gxa,Cai:2017mow,Dev:2018sel}.}

\yong{We focus on the type-II seesaw model in this work, which is obtained by extending the SM Higgs sector with a complex triplet that transforms as (1,3,2) under the SM gauge group. The type-II seesaw model differents from the other two seesaw models in that it allows large neutrino Yukawa couplings simultaneously with a light seesaw scale even below TeV. This can be realized with a small triplet vacuum expectation value that naturally generates tiny neutrino masses with even $\mathcal{O}(1)$ neutrino Yukawa couplings\,\cite{Du:2018eaw}. In addition, since the complex scalar transforms as a triplet under $\rm SU(2)_L$, new interactions between the SM Higgs doublet and the complex triplet will present and modify the Higgs potential.\footnote{See also Refs.~\cite{Niemi:2018asa,Zhou:2018zli,Bian:2019bsn,Addazi:2019dqt,Niemi:2020hto} for a similar work on electroweak phase transition in different scenarios.} The modified Higgs potential could then change the phase transition type of the SM, thus also serve as a possible candidate for explaining the observed baryon asymmetry of the Universe\,\cite{Planck:2018vyg}.\footnote{While the complex triplet is feasible to simultaneously explain the baryon asymmetry and non-vanishing neutrino masses, it does not provide any dark matter candidate since experimental results prohibit the neutral component of a complex triplet with $-2$ hyperchange from being both light and stable\,\cite{OPAL:1992glr,OPAL:2001luy}. This can be circumvented by a real triplet with vanishing hypercharge, where the triplet around $\sim250$\,GeV is still allowed from the disappearing track searches\,\cite{Chiang:2020rcv}, and its connection to the baryon asymmetry can be found in\,\cite{Bell:2020gug}. Alternatively, dark matter and the baryon asymmetry could be simultaneously explained by adding a dark sector to the complex triplet, where the baryon asymmetry is realized through lepton asymmetry conversion in the dark sector, see\,\cite{Hall:2021zsk} for the details.} While this model have been intensively studied experimentally\,\cite{OPAL:1992glr,OPAL:2001luy,Aaltonen:2011rta,Aad:2012cg,ATLAS:2012mn,ATLAS:2012hi,Aad:2014hja,ATLAS:2014kca,Aad:2015oga,Sirunyan:2017ret,Aaboud:2017qph} and theoretically\,\cite{Machacek:1983tz,Machacek:1983fi,Machacek:1984zw,Arason:1991ic,Ford:1992pn,Barger:1992ac,Luo:2002ey,Chao:2006ye,Schmidt:2007nq,Dey:2008jm,Arhrib:2011uy,Chao:2012mx,Chun:2012jw,Bonilla:2015eha,Haba:2016zbu,Cai:2017mow,Li:2018jns,Agrawal:2018pci,Du:2018eaw} at colliders, the pattern of its phase transition in this model have not yet been investigated to the best of our knowledge.}

\yong{As mentioned in last paragraph, the modified Higgs potential, due to new interactions between the doublet and the triplet, could change the phase transition of the SM Higgs from a crossover type to a strong first-order phase transition. The strong first-order phase transition is a necessary condition that validates the departure from thermal equilibrium, one of the three Sakharov's conditions~\cite{Sakharov:1967dj}. As a result, the triplet would be possible to explain the dynamic generation of the baryon asymmetry of the Universe through the electroweak baryogenesis paradigm\,\cite{Kuzmin:1985mm,Cohen:1990it,Cohen:1993nk,Quiros:1994dr,Rubakov:1996vz,Funakubo:1996dw,Trodden:1998ym,Bernreuther:2002uj,Morrissey:2012db,DiBari:2013rga}. On the other hand, stochastic background of gravitational waves could also be generated during the first-order phase transition. And recently, the observation of gravitational waves from LIGO and VIRGO has opened a new window to probe new physics beyond the SM\,\cite{LIGOScientific:2016aoc,LIGOScientific:2017vwq,LIGOScientific:2018mvr,LIGOScientific:2020ibl} -- For a comprehensive discussion on this point, see, for example, Refs.\,\cite{Mazumdar:2018dfl,Caprini:2019egz} and references therein. Therefore, it would be interesting to investigate the role that can be played by current and future gravitational wave observatories, such as LISA\,\cite{LISA:2017pwj}, TianQin\,\cite{TianQin:2015yph, Hu:2018yqb, TianQin:2020hid}, Taiji\,\cite{Hu:2017mde, Ruan:2018tsw}, DECIGO\,\cite{Seto:2001qf, Kudoh:2005as}, and BBO\,{\cite{Ungarelli:2005qb, Cutler:2005qq}}\footnote{The AION/MAGIS and AEDGE would able to probe the mid frequency band~\cite{Badurina:2019hst,AEDGE:2019nxb,Badurina:2021rgt}.}, in searching for new physics models like the type-II seesaw model considered in this work,\footnote{Recently, there are various studies on how to probe the seesaw scale of type-I or type-I like seesaw models with gravitational waves from phase transition\,\cite{Brdar:2019fur,Brdar:2018num,Okada:2018xdh,Bian:2019szo,Li:2020eun,Costa:2022oaa} and cosmic strings~\cite{Dror:2019syi,Bian:2021vmi,Blasi:2020wpy}.} and also possibly its complementarity with collider searches or other low-energy precision experiments.}

\yong{The phase transition pattern of the complex triplet model is studied in detail in this work, based on which we then study the generated gravitational waves from the transition and their observation at current and future gravitational wave observatories mentioned above. The rest of this work is organized as follows. In section\,\ref{sec:modelsetup}, we briefly review the type-II seesaw model and the model constraints. Then in section\,\ref{sec:phasetrans}, we calculate the pattern of phase transition in this model and obtain possible benchmark points for a strong first-order electroweak phase transition. Section\,\ref{sec:gravwave} is devoted to the study of gravitational wave production from the phase transition, and we then conclude in section\,\ref{sec:conclu}.}

\section{The model}\label{sec:modelsetup}

As discussed in the introduction, the type-II seesaw model can naturally induce non-vanishing neutrino masses that are responsible for neutrino oscillations. In addition, the type-II seesaw model also introduces new interactions for the Higgs doublet, which could distort the SM Higgs potential and thus possibly permit a first-order phase transition. In this section, we will firstly briefly review the details of this model and then discuss its theoretical constraints.

\subsection{Model setup}\label{model:setup}

The complex triplet Higgs model (CTHM) can be obtained by extending the SM Higgs portal with a complex triplet $\Delta$ that transforms as $(1,3,2)$ under that SM gauge group. The Lagrangian of this model can be written as
\eqal{\mathcal{L}_{\rm CTHM}=\mathcal{L}_{\rm SM} + \mathcal{L}_{\rm kinetic} - V_{\rm CTHM},}
with the kinetic part and the most general form of the CTHM potential given as, respectively,
\begin{align}
\mathcal{L}_{\rm{kin}}&={\rm{Tr}}[(D_\mu \Delta)^\dagger (D^\mu \Delta)],\quad \text{where } D_\mu \Delta\equiv\partial_\mu \Delta + \frac{ig}{2}[\tau^aW_\mu^a,\Delta]+\frac{ig'Y_{\Delta}}{2}B_\mu\Delta,\\
V(\Phi,\Delta)&= - m^2\Phi^\dagger\Phi + M^2{\rm{Tr}}(\Delta^\dagger\Delta)+\left[\mu \Phi^Ti\tau_2\Delta^\dagger \Phi+\rm{h.c.}\right]+\lambda_1(\Phi^\dagger\Phi)^2 \nonumber\\
&~~~~+\lambda_2\left[\rm{Tr}(\Delta^\dagger\Delta)\right]^2 +\lambda_3\rm{Tr}[ \Delta^\dagger\Delta \Delta^\dagger\Delta]
+\lambda_4(\Phi^\dagger\Phi)\rm{Tr}(\Delta^\dagger\Delta)+\lambda_5\Phi^\dagger\Delta\Delta^\dagger\Phi.\label{eq:tripletpotential}
\end{align}

Note that the kinetic term introduces new interactions between $Z$, $W^\pm$ and the triplet $\Delta$. As a result, when the triplet gets a non-vanishing vacuum expectation value (vev) after electroweak spontaneous symmetry breaking, the SU(2) gauge boson masses will receive non-zero corrections from the triplet. \yong{Note also that the $\mu$ term in the Lagrangian explicitly violates lepton numbers by two units, such that $\mu^2/M^2$ can be effectively used to efficiently estimate the extent to which lepton number will be violated.} On the other hand, the $\lambda_{4,5}$ terms in the potential induce new interactions to the Higgs doublet such that the Higgs potential would be distorted during the evolution of the Universe, making this model also a possible candidate for explaining the observed baryon number asymmetry of the Universe (BAU) through electroweak baryogenesis. While this possibility has been pointed out, for example, in Ref.\,\cite{Du:2018eaw}, the authors only focused on collider studies of this model. In this work, we extend their investigation to include a detailed study on the electroweak phase transition and also \yong{on the generated gravitational waves from considering both current and future gravitational wave experiments}. We postpone our discussion on this point to sections\,\ref{sec:phasetrans} and \ref{sec:gravwave}, and focus on the broken scenario of this model for the moment in the following.

After electroweak spontaneous symmetry breaking, we parameterize the SM Higgs $\Phi$ and the triplet $\Delta$ in the following forms:
\begin{eqnarray}\label{basis}
\Phi=\left(
\begin{array}{c}
\varphi^+\\
\frac{1}{\sqrt{2}}(\varphi+v_\Phi+i\chi)
\end{array}\right), \quad
\Delta =
\left(
\begin{array}{cc}
\frac{\Delta^+}{\sqrt{2}} & H^{++}\\
\frac{1}{\sqrt{2}}(\delta+v_\Delta+i\eta) & -\frac{\Delta^+}{\sqrt{2}}
\end{array}\right).
\end{eqnarray}
where $v_\Delta$ ($v_\Phi$) is the vev of the triplet (doublet). The neutrino masses can then be generated through the following Yukawa Lagrangian:
\eqal{\mathcal{L}_Y= & (y_\nu)_{\alpha\beta}\overline{L^{ c}_\alpha}i\tau_2\Delta L_\beta+\rm{h.c.},\label{nvlag}}
where $\alpha$ and $\beta$ are the lepton flavor indices and $\tau_2$ is the second Pauli matrix. The neutrino mass matrix can be expressed as\footnote{After integrating out the triplet, this Yukawa Lagrangian naturally generates the dimension-5 Weinberg operator. The full tree- and one-loop matching between this model and the SMEFT is presented recently in Refs.\,\cite{Du:2022vso,Li:2022ipc}.}
\eqal{(m_{\nu})_{\alpha\beta} = \sqrt{2}(y_\nu)_{\alpha\beta}v_\Delta.}
Due to the smallness of neutrino masses\,\cite{ParticleDataGroup:2020ssz}, the neutrino Yukawa couplings $(y_\nu)_{\alpha\beta}$ would be constrained to be very tiny for $v_\Delta\sim\mathcal{O}(1\rm\,GeV)$. Similarly, for $(y_\nu)_{\alpha\beta}\sim\mathcal{O}(1)$, the triplet vev $v_\Delta$ would also be required to be tiny.

On the other hand, a non-vanishing $v_\Delta$ would also introduce mixing between the SM Higgs and the triplet through the $\lambda_{4,5}$ terms in the potential. As a consequence, the Higgs particles are not in their mass eigenstates. Following the notations established in Ref.\,\cite{Du:2018eaw}, we define
\begin{eqnarray}
\left(\begin{array}{c}h\\H\\G^0\\ A\\G^\pm\\H^\pm \end{array}\right)&=&\left(\begin{array}{cccccc}
\cos \alpha & \sin\alpha & 0 & 0 & 0 & 0\\ -\sin\alpha   & \cos\alpha & 0 & 0 & 0 & 0\\
0 & 0 & \cos \beta_0 & \sin\beta_0 & 0 & 0\\ 0 & 0 & -\sin\beta_0 & \cos\beta_0 & 0 & 0\\
0 & 0 & 0 & 0 & \cos \beta_\pm & \sin\beta_\pm \\ 0 & 0 & 0 & 0 & -\sin\beta_\pm & \cos\beta_\pm\end{array}\right)\left(\begin{array}{c} \varphi\\\delta\\\chi\\\eta\\\varphi^\pm\\\Delta^\pm\end{array}\right), \label{rotation matrix}
\end{eqnarray}
with $h,\,H,\,G^0,\, A,\,G^\pm,\,H^\pm$ being the mass eigenstates and the mixing angles being
\begin{eqnarray}
\tan\beta_\pm=\frac{\sqrt{2}v_\Delta}{v_\Phi}, \quad \tan\beta_0=\frac{2v_\Delta}{v_\Phi}, \quad \tan2\alpha &=&\frac{v_\Delta}{v_\Phi}\cdot\frac{2v_\Phi  \lambda_{45}-\frac{2 \sqrt2 \mu v_\Phi}{v_\Delta} }{2v_\Phi\lambda_1-\frac{v_\Phi\mu}{\sqrt{2}v_\Delta}-\frac{2 v_\Delta^2 \lambda_{23}}{v_\Phi}}. \label{tan2a}
\end{eqnarray}

The mass eigenvalues can then be expressed as a function of the mixing angles and the model parameters\,\cite{Du:2018eaw}:
\begin{eqnarray}
&&M_{H^{\pm\pm}}^2=M_\Delta^2-v_\Delta^2\lambda_3-\frac{\lambda_5}{2}v_\Phi^2,\label{mhpp}\\
&&M_{H^\pm}^2=\left(M_\Delta^2-\frac{\lambda_5}{4}v_\Phi^2\right)\left(1+\frac{2v_\Delta^2}{v_\Phi^2}\right),\label{mhp}
\end{eqnarray}
\begin{eqnarray}
&&M_A^2 =M_\Delta^2\left(1+\frac{4v_\Delta^2}{v_\Phi^2}\right), \label{mA}\\
&&M_h^2= 2v_\Phi^2\lambda_1\cos^2\alpha+\left( M_\Delta^2+2\lambda_{23}v_\Delta^2\right) \sin^2\alpha + \left( \lambda_{45} v_\Phi v_\Delta - \frac{2v_\Delta}{v_\Phi}M_\Delta^2\right) \sin2\alpha,\label{mh}\\
&&M_H^2=2v_\Phi^2\lambda_1\sin^2\alpha+ \left( M_\Delta^2+2\lambda_{23}v_\Delta^2 \right) \cos^2\alpha -  \left( \lambda_{45} v_\Phi v_\Delta - \frac{2v_\Delta}{v_\Phi}M_\Delta^2 \right) \sin2\alpha,\label{mH}
\end{eqnarray}
with
\eqal{
M_\Delta^2\equiv&\, \frac{v_\Phi^2\mu}{\sqrt{2}v_\Delta},\quad \lambda_{ij}\equiv\, \lambda_i + \lambda_j.
}
One key observation from the expressions above is that $\lambda_{2,3}$ always appear in pair with $v_\Delta^2$. This can be easily understood from the fact that two of the four triplets have to take their corresponding vevs to contribute to the mass terms. However, as we shall see shortly below, precision measurement of the $\rho$ parameter requires $v_\Delta$ to be small, thus suppressing any observable effects from $\lambda_{2,3}$ phenomenologically. For this reason, we fix $\lambda_2 = 0.2$ and $\lambda_3=0$ for our study below and comment again on the fact that different values of $\lambda_{2,3}$ barely have any impact on our conclusions below.

\subsection{Model constraints}\label{model:const}
As mentioned in last subsection, the triplet model would modify the SU(2) gauge boson masses through the kinetic part of the Lagrangian. The corrections, however, could not be too large to be consistent with experimental results. In this section, we briefly summarize constraints from the $\rho$ parameter\,\cite{ParticleDataGroup:2020ssz}, LHC constraints\,\cite{ATLAS:2017xqs} on the mass scale of the triplet, and theoretical constraints from vacuum stability, perturbative unitarity and purturbativity\,\cite{Arhrib:2011uy,Haba:2016zbu,Chao:2006ye,Schmidt:2007nq, Bonilla:2015eha,Machacek:1983tz,Machacek:1983fi,Machacek:1984zw,Ford:1992pn,Arason:1991ic,Barger:1992ac,Luo:2002ey,Chao:2012mx,Chun:2012jw}.

\subsubsection{Constraints from the $\rho$ parameter}
The $\rho$ parameter is defined
\eqal{\rho\equiv\frac{M_W^2}{M_Z^2\cos^2\theta_W},}
where $M_W$ ($M_Z$) is the mass of $W^\pm$ ($Z$) and $\theta_W$ is the weak mixing angle. \yong{After electroweak spontaneous symmetry breaking, the triplet invents non-vanishing corrections to $M_{W,Z}$ through the kinetic Lagrangian. At tree level, the $\rho$ parameter can then be expressed as}
\begin{eqnarray}
\rho=\frac{v_\Phi^2+2v_\Delta^2}{v_\Phi^2+4v_\Delta^2}\approx1-\frac{2v_\Delta^2}{v_\phi^2}.
\end{eqnarray}
In the case where the triplet does not develop a non-vanishing vev, one reproduces the tree-level SM prediction of $\rho=1$. Experimentally, the $\rho$ parameter has been \yong{measured} to be $\rho=1.00038\pm0.00020$\,\cite{ParticleDataGroup:2020ssz}, resulting in
\begin{eqnarray}
0\le v_\Delta \lesssim 2.56 {\rm ~\,GeV.}
\end{eqnarray}
Note that since $v\equiv\sqrt{v_\Phi^2+v_\Delta^2}=\left(\sqrt{2}G_F\right)^{-1/2}\approx246\rm\,GeV$ with $G_F$ the Fermi constant determined from the muon lifetime, one immediately concludes that $v_\Delta\ll v_\Phi$.

\subsubsection{Theoretical constraints}\label{subsubsec:thcons}
Theoretical constraints on the triplet model have been well documented in literature, we summarize these constraints below based on Refs.\,\cite{Arhrib:2011uy,Haba:2016zbu,Chao:2006ye,Schmidt:2007nq,Dey:2008jm,Bonilla:2015eha,Machacek:1983tz,Machacek:1983fi,Machacek:1984zw,Ford:1992pn,Arason:1991ic,Barger:1992ac,Luo:2002ey,Chao:2012mx,Chun:2012jw}. Specifically, we comment on that perturbativity has been found to put very stringent constraints on the model parameter space. For this reason, we include perturbativity up to one-loop in this work and point out that two-loop results for the portal couplings have been studied in Ref.\,\cite{Chao:2012mx}.
\begin{itemize}
\item Vacuum stability:
\begin{align}
\lambda_1\ge0,\quad \lambda_2&+\text{min}\left\{\lambda_3,\frac{\lambda_3}{2}\right\}\ge0, \nonumber\\
\lambda_4+\text{min}\left\{0,\lambda_5\right\}& + \text{min}\left\{2\sqrt{\lambda_1\lambda_{23}},2\sqrt{\lambda_1(\lambda_2+\frac{\lambda_3}{2})}\right\}\ge0.
\end{align}

\item Perturbative unitarity:

\begin{align}
&|\lambda_1|\le4\pi,  \quad|\lambda_2|\le4\pi, \quad |\lambda_{23}|\le4\pi, \nonumber\\
&|\lambda_4 - \frac{\lambda_5}{2} |\le8\pi, \quad |2\lambda_2-\lambda_3|\le8\pi, \nonumber\\
&|\lambda_{45}|\le8\pi, \quad |\lambda_4|\le8\pi, \quad |2\lambda_4+3\lambda_5|\le16\pi,\nonumber\\
&|\lambda_{12}+2\lambda_3 \pm\sqrt{(\lambda_1-\lambda_2-2\lambda_3)^2+\lambda_5^2}|\le8\pi, \nonumber\\
&|3\lambda_{13}+4\lambda_2 \pm\sqrt{(3\lambda_1-4\lambda_2-3\lambda_3)^2+\frac{3}{2}(2\lambda_4+\lambda_5)^2}|\le8\pi,\label{peruni}
\end{align}

\item Perturbativity:
\begin{align}
\left(4\pi\right)^{2}\frac{dg_{i}}{dt} & =b_{i}g_{i}^{3}\textrm{ with }b_{i}=\left(\frac{47}{10},-\frac{5}{2},-7\right)\,,\\
\left(4\pi\right)^{2}\frac{dy_t}{dt} & =y_t\left[\frac{9}{2}y_t^2-\left(\frac{17}{20}g_1^2+\frac{9}{4}g_2^2+8g_3^2\right)\right]\,,\\
\left(4\pi\right)^{2}\frac{d\lambda_{1}}{dt} & =\frac{27}{200}g_{1}^{4}+\frac{9}{20}g_{1}^{2}g_{2}^{2}+\frac{9}{8}g_{2}^{4}-\left(\frac{9}{5}g_{1}^{2}+9g_{2}^{2}\right)\lambda_{1}+24\lambda_{1}^{2}+3\lambda_{4}^{2}\nonumber \\
 & + 3\lambda_{4}\lambda_{5} + \frac{5}{4}{\lambda_{5}}^{2} + 12\lambda_{1}y_{t}^{2}-6y_{t}^{4}\,,\\
 \left(4\pi\right)^{2}\frac{d\lambda_{2}}{dt} & =\frac{54}{25}g_{1}^{4}-\frac{36}{5}g_{1}^{2}g_{2}^{2}+15g_{2}^{4}-\left(\frac{36}{5}g_{1}^{2}+24g_{2}^{2}\right)\lambda_{2}+2\lambda_{4}^{2}+2\lambda_{4}\lambda_{5}\nonumber \\
 & +28\lambda_{2}^{2}+24\lambda_{2}\lambda_{3}+6{\lambda_{3}}^{2}\,,\\
\left(4\pi\right)^{2}\frac{d\lambda_{3}}{dt} & =\frac{72}{5}g_{1}^{2}g_{2}^{2}-6g_{2}^{4}+{\lambda_{5}}^{2}-\left(\frac{36}{5}g_{1}^{2}+24g_{2}^{2}\right)\lambda_{3}+24\lambda_{2}\lambda_{3}+18{\lambda_{3}}^{2}\,,\\
\left(4\pi\right)^{2}\frac{d\lambda_{4}}{dt} & =\frac{27}{25}g_{1}^{4}-\frac{18}{5}g_{1}^{2}g_{2}^{2}+6g_{2}^{4}-\left(\frac{9}{2}g_{1}^{2}+\frac{33}{2}g_{2}^{2}\right)\lambda_{4}+12\lambda_{1}\lambda_{4}+4\lambda_{1}\lambda_{5}\nonumber \\
 & +4\lambda_{4}^{2} +16\lambda_{2}\lambda_{4}+12\lambda_{3}\lambda_{4}+{\lambda_{5}}^{2}+6\lambda_{2}\lambda_{5}+2\lambda_{3}\lambda_{5}+6\lambda_{4}y_{t}^{2}\,,
\end{align}
\begin{align}
\left(4\pi\right)^{2}\frac{d\lambda_{5}}{dt} & =\frac{36}{5}g_{1}^{2}g_{2}^{2}-\left(\frac{9}{2}g_{1}^{2}+\frac{33}{2}g_{2}^{2}\right)\lambda_{5}+4\lambda_{1}\lambda_{5}+8\lambda_{4}\lambda_{5}+4{\lambda_{5}}^{2}+4\lambda_{2}\lambda_{5}\nonumber \\
 & +8\lambda_{3}\lambda_{5}+6\lambda_{5}y_{t}^{2}\,.
\end{align}
with $t\equiv\ln(\mu/M_t)$, $\mu$ the 't Hooft scale, $y_t$ the top Yukawa, and $M_t=173.1$\,GeV being our input scale. \yong{All other input SM parameters at this scale are taken from Ref.\,\cite{Buttazzo:2013uya}.}
\end{itemize}

\subsubsection{Collider constraints}
The smoking-gun signature of the triplet model is the same-sign dilepton final state from the decay of $H^{\pm\pm}$. The same-sign dilepton channel has \yong{an almost} 100\% branching ratio when the triplet vev is small, or equivalently when the neutrino Yukawa $(y_\nu)_{\alpha\beta}$ is of $\mathcal{O}(1)$\,\cite{Du:2018eaw}. The ATLAS collaboration\,\cite{ATLAS:2017xqs} reported the most stringent constraint on $M_{H^{\pm\pm}}$ in this case from the same-sign di-muon final state, which is

\eqal{M_{H^{\pm\pm}}\gtrsim870\rm\,GeV\quad\text{( Assuming Br($H^{\pm\pm}\to\mu^\pm\mu^\pm$)=100\% )}.}

We comment on that the lower bound on the triplet scale above is only valid when $v_\Delta$ is large or equivalently when the neutrino Yukawa couplings are tiny of order $m_\nu/{\rm GeV}$. However, for $y_\nu$ of $\mathcal{O}(1)$, the same-sign dilepton final state will be highly suppressed and the same-sign di-$W$ boson would dominate instead\,\cite{Du:2018eaw}. For the recent report from the ATLAS collaboration on the same-sign vector boson final states, see\,\cite{ATLAS:2021jol}, and we comment on that the lower bound on the triplet mass in this case is then much weaker than the one above.

As we shall see in section\,\ref{sec:phasetrans}, a relatively light triplet helps trigger a strong first-order electroweak phase transition (SFOEWPT) that could be responsible for the BAU as well as the production of gravitational waves, both of which barely have any sensitivity to the value of $v_\Delta$. Therefore, a relatively light triplet with small $v_\Delta$ would be the promising scenario for a SFOEWPT and the generation of gravitational waves. Furthermore, a small $v_\Delta$ also implies an $\mathcal{O}(1)$ neutrino Yukawa couplings, making the seesaw Lagrangian more natural. The detail of our analysis for drawing the conclusions above on the phase transition and the gravitational waves will be detailed in the next two sections.

\section{Electroweak phase transition in the triplet model}\label{sec:phasetrans}
The 125\,GeV Higgs particle observed at the LHC\,\cite{ATLAS:2012yve,CMS:2012qbp} suggests the phase transition in the SM is of a crossover type\,\cite{Kajantie:1995kf,Kajantie:1996mn,Kajantie:1996qd}. As a result, the SM is short of explaining the observed BAU through electroweak baryogenesis since the latter requires a SFOEWPT. Due to the presence of the complex triplet, the Higgs potential would be modified by the $\lambda_{4,5}$ terms in eq.\,\eqref{eq:tripletpotential}, which \yong{introduce} extra interactions between the doublet and the triplet. Therefore, proper values of $\lambda_{4,5}$ could modify the Higgs potential in a way such that a SFOEWPT could be realized. This would be the topic of this section.

To that end, we start from the scalar potential at finite temperatures and parameterize the effective potential $V_{\rm eff} (\phi,\delta,T)$ generically as
\begin{align}
\label{potVeff}
V_{\rm eff}(\phi,\delta,T) &= V_0(\phi,\delta) +V_{\rm CW}(\phi,\delta) +V_{\rm CT}(\phi,\delta) + V_{\rm th}(\phi,\delta,T) + V_{\rm daisy}(\phi,\delta,T)\; ,
\end{align}
where $V_0(\phi,\delta)$ is the tree-level potential, $V_{\rm CW}(\phi,\delta)$ is the Coleman-Weinberg potential, $V_{\rm CT}$ is the counter-term (CT) corrections fixed by fulfilling the tree-level relations of the parameters in $V_0$, $V_{\rm th}(\phi,\delta,T)$ and $V_{\rm daisy}$ are the leading thermal corrections.

The pattern of phase transition in specific UV models depends on correctly accounting for each part in eq.\,\eqref{potVeff}. For this reason, we review the results term by term in the following subsections.
\subsection{The tree level potential }
The tree-level potential will be a function of the complex doublet and the complex triplet fields. To simplify the calculation, one can remove the Goldstone modes by properly performing $\rm SU(2)$ gauge transformations\,\cite{Cline:1996mga,Cline:2011mm}. It then suffices to focus on the neutral components of this model, which can be readily obtained as
\begin{align}
\label{Vtree}
V_0(\phi,\delta) = &\, \frac{\lambda_1}{4} (\phi^4-2 v_\phi^2 \phi^2)+\frac{\lambda_{23}}{4} (\delta^4-2 v_\Delta^2 \delta ^2) \nonumber\\
& + \frac{\lambda_{45}}{4} (\phi ^2 (\delta ^2-v_\Delta^2)-\delta ^2 v_\phi^2)+\frac{\mu \phi^2(v_\Delta-\delta)}{\sqrt{2}v_\Delta} + \frac{1}{2} m_\Delta^2 \delta^2 .
\end{align}

\subsection{The Coleman-Weinberg potential }
It is well known that loop corrections could change the pattern of electroweak symmetry breaking, see\,\cite{Coleman:1973jx}.\footnote{Recently, this complex triplet model has been investigated in Ref.\,\cite{Du:2022vso} at zero temperature for radiative symmetry breaking at one loop.} Systematically, the zero temperature effective potential, referred to as the Coleman-Weinberg (CW) potential in the following, could be derived following the procedure outlined in\,\cite{Coleman:1973jx}. Using the $\overline{\rm MS}$ scheme and taking the Landau gauge to decouple any ghost contributions, one can generically write the one-loop CW potential in the following form\,\cite{Quiros:2003gg}:
\begin{align}
V_{\rm CW}(\phi,\delta) = \sum_{i} (-1)^{2s_i} n_i\frac{M_i^4(\phi,\delta)}{64\pi^2}\left[\ln \frac{M_i^2 (\phi,\delta)}{\mu^2}-C_i\right] \; \ ,
\end{align}
where the sum $i$ runs over contributions from all particles in the theory, $s_i$ and $n_i$ are the spin and the number of degrees of freedom, respectively, with $n_{h,H,A,H^\pm,H^{\pm\pm},G^0,G^\pm,W^\pm,Z,t}={1,1,1,2,2,1,2,6,3,12}$. $\mu$ is the renormalization scale for which we fix at $\mu=v$, and $C_i$ are renormalization scheme dependent constants. In this work, we adopt the $\overline{\text{MS}}$ on-shell scheme with $C_{W^\pm,Z}=5/6$ and $C_{i}=3/2$ otherwise.

\subsection{The counter-term potential }
As originally noticed in\,\cite{Coleman:1973jx}, inclusion of $V_{\rm CW}$ will shift the minimum of the Higgs potential at tree level. As a result, the minimization conditions of the tree Lagrangian no longer hold. The CT potential could thus be added to restore these tree-level relations from our renormalization conditions just discussed above. To be specific, upon parameterizing the CT potential as
\begin{align}
V_{\rm CT}= \delta m^2 \phi^2+\delta M^2 \delta^2 +\delta \lambda_1 \phi^4+\delta \lambda_{23} \delta^4 +\delta \lambda_{45} \phi^2\delta^2 \; ,
\end{align}
one can readily solve these CTs from the following minimization conditions:
\begin{align}
&\frac{\partial V_{\rm CT}}{\partial \phi}+\frac{\partial V_{\rm CW}}{\partial \phi}=0\;, \quad \frac{\partial V_{\rm CT}}{\partial \delta}+\frac{\partial V_{\rm CW}}{\partial \delta}=0\;,\label{eq:ct1}\\
& \frac{\partial^{2} V_{\rm CT}}{\partial \phi\partial\delta}+\frac{\partial^{2} V_{\rm CW}}{\partial \phi\partial\delta}=0\;,\quad \frac{\partial^{2} V_{\rm CT}}{\partial \phi^2}+\frac{\partial^{2} V_{\rm CW}}{\partial \phi^2}=0\;, \quad \frac{\partial^{2} V_{\rm CT}}{\partial \delta^2}+\frac{\partial^{2} V_{\rm CW}}{\partial \delta^2}=0\;. \label{eq:ct2}
\end{align}

One immediate problem, however, arises for the Goldstone bosons when solving the CTs from conditions above and the reason is as follows. Since we work in the Landau gauge to decouple the ghosts from $V_{\rm CW}$, the Goldstone bosons become massless under this specific choice of gauge. As a result, when one calculates the CTs from above conditions, terms of $(\partial^2M_{\rm G.B.}/\partial\phi^2)\times\log(M_{\rm G.B.}^2)$ and/or $(\partial^2M_{\rm G.B.}/\partial\delta^2)\times\log(M_{\rm G.B.}^2)$ with $M_{\rm G.B.}$ the Goldstone boson masses, will be generated with non-vanishing prefactors ahead of $\log(M_{\rm G.B.}^2)$. Thus, the logarithmic divergence from vanishing $M_{\rm G.B.}$ renders the Higgs masses renormalized at vanishing momentum from Goldstone particles ill-defined. To circumvent this issue, we follow the strategy in\,\cite{Cline:2011mm} by introducing an infrared cutoff scale $m_{\rm IR}$ at $m_{\rm IR}=m_h$ and replacing $M_{\rm G.B.}$ by $m_{\rm IR}$ in eqs.\,\eqref{eq:ct1}-\eqref{eq:ct2}. We comment on that a more exact solution for this issue can be found in \,\cite{Cline:1996mga}, and our approach produces consistent results when adopting the more exact method.

\subsection{The thermal effective potential }

The finite temperature corrections to the effective potential at one-loop can be obtained from calculating the free energy of bosonic and fermionic particles that obtain masses from $\phi$ and $\delta$, which can be expressed as~\cite{Dolan:1973qd}
\begin{align}
\label{potVth}
 V_{\rm th}(\phi,\delta, T) = \frac{T^4}{2\pi^2}\, \sum_i n_i J_{B,F}\left( \frac{ M_i^2(\phi,\delta)}{T^2}\right)\;,
\end{align}
where $n_{B,F}$ are the numbers of degrees of freedom for bosonic and fermionic particles, respectively. $J_{B(F)}$ are the thermal integrals for bosonic (fermionic) particles defined as
\eqal{J_{B(F)} = \pm \int_0^\infty dx x^2\ln\left( 1\mp e^{-\sqrt{x^2+\beta^2m_{B(F)}^2}} \right),
}
with $\beta\equiv1/T$ and the upper (lower) sign for bosonic (fermionic) particles. Numerically, above expressions can be efficiently calculated by expanding $J_{B(F)}$ in terms of the modified Bessel functions of the second kind $K_{2} (x)$\cite{Anderson:1991zb}:
\begin{align}
\label{Bessel_JFJB}
J_{B,F}(y) = \lim_{N \to +\infty} \mp \sum_{l=1}^{N} {(\pm1)^{l}  y \over l^{2}} K_{2} (\sqrt{y} l),
\end{align}
with $y\equiv m_{i}^2(\phi,\delta)/T^2$ and the upper (lower) sign corresponds to bosonic (fermionic) contributions.

Finally, there is another important part of the thermal corrections to the scalar masses coming from the resummation of \textit{ring }(or\textit{\ daisy})
diagrams~\cite{Carrington:1991hz,Arnold:1992rz}\footnote{See Refs.\cite{Croon:2020cgk,Schicho:2022wty,Schicho:2021gca,Niemi:2021qvp} for the effective theory constructed using Dimensional Reduction, which established the method to systematically incorporate thermal contributions to the masses and couplings.},
\begin{align}
V_{\rm daisy}\left(\phi,\delta,T\right) =-\frac{T}{12\pi }\sum_{i} n_{i}\left[ \left( M_{i}^{2}\left(\phi,\delta,T\right) \right)^{\frac{3}{2}}-\left( M_{i}^{2}\left( \phi,\delta\right) \right)^{\frac{3}{2}}\right] ,
\label{eq:daisy}
\end{align}
where $M_{i}^{2}\left( \phi,\delta,T\right) $ are the thermal Debye masses of the
bosons corresponding to the eigenvalues of the full mass matrix
\begin{align}
\label{eq:thermalmass}
M_{i}^{2}\left( \phi,\delta,T\right) ={\rm eigenvalues} \left[ \hat{m}_{X}^{2}\left( \phi,\delta\right) +\Pi^{X}(T)\right]  ,
\end{align}
which consists of the field dependent mass matrices at $T=0$:

\begin{align}
\label{eq:fieldmassP}
\hat{m}^2_{P} & = \begin{pmatrix}
-m^2+\frac{1}{2}\lambda_{45}\delta^2-\sqrt{2}\mu\delta+3\lambda_1 \phi^2&  \quad  \lambda_{45}\delta\phi-\sqrt{2}\mu\phi \\[5pt]
\lambda_{45}\delta\phi-\sqrt{2}\mu\phi &  \quad M^2+3\lambda_{23}\delta^2 + \frac{1}{2}\lambda_{45} \phi^2
\end{pmatrix}, \\
\label{eq:fieldmassA}
\hat{m}^2_{A} & = \begin{pmatrix}
-m^2+\frac{1}{2}\lambda_{45}\delta^2+\sqrt{2}\mu\delta+\lambda_1 \phi^2 &  \quad -\sqrt{2}\mu\phi  \\[5pt]
-\sqrt{2}\mu\phi &  M^2+\lambda_{23}\delta^2 + \frac{1}{2}\lambda_{45} \phi^2
\end{pmatrix}, \\
\label{eq:fieldmasspm}
\hat{m}^2_{\pm} & = \begin{pmatrix}
-m^2+\lambda_1 \phi^2+\frac{\delta^2 \lambda_4}{2} &  \quad \frac{\sqrt{2}}{4}\lambda_5 \delta \phi-\mu\phi \\[5pt]
\frac{\sqrt{2}}{4}\lambda_5 \delta \phi-\mu\phi &  \quad
M^2+\lambda_{23}\delta^2+\frac{1}{4}(2\lambda_4+\lambda_5)\phi^2
\end{pmatrix}, \\
\label{eq:fieldmasspmpm}
\hat{m}^2_{\pm\pm} & = M^2 + \lambda_2 \delta^2 + \frac{1}{2}\lambda_4 \phi^2,
\end{align}
and the finite temperature corrections of $\Pi^X(T)\, (X=P,A,\pm,\pm\pm)$:
\begin{align}
\label{eq:thermalcorrection}
\Pi^{P/A/\pm}(T) & = \begin{pmatrix}
\Pi^{P/A/\pm}_{11}(T) &  \quad \Pi^{P/A/\pm}_{12}(T)  \\[5pt]
\Pi^{P/A/\pm}_{12}(T) &  \quad \Pi^{P/A/\pm}_{22}(T)
\end{pmatrix},
\end{align}
with {\color{black}the non-diagonal elements being zero and} the diagonal elements being
\begin{align}
\label{eq:thermalmass2}
\Pi^{P/A/\pm}_{11}(T) &= \frac{1}{16} T^2 (3 g^2+g'^2+2 \lambda_5+4 (2 \lambda_1 +\lambda _4+y_t^2))\;, \nonumber\\
\Pi^{\pm\pm}(T) &= \Pi^{P/A/\pm}_{22}(T) = \frac{1}{12} T^2 (6 g^2+3 g'^2+{\color{black}8} \lambda_2+6\lambda_3+2\lambda_4+\lambda_5)\;.
\end{align}
{\color{black}Then with the help of rotation matrix defined in eq.\,\eqref{rotation matrix}, one can readily obtain the corresponding mass eigenstates.}

{\color{black}With the effective potential at one loop fully determined, one can then investigate the patterns of phase transition. \yong{In particular, when a potential barrier presents between the false and the true vacua at the critical temperature, a first-order phase transition would occur.
Furthermore,} to ensure the coexistence of degenerated vacua at the critical temperature $T_c$, we use the determinant of the finite-temperature Hessian matrix together with the following conditions:}
\eqal{M_{3}P_{3} - {N_{3}^2} > 0,M_{3} > 0,}
where
\begin{align}\label{treehess}
M_3 &\equiv \left.\frac{{{d^2}V_{\rm eff}({\phi },{\delta },{T_{c}})}}{{d\phi ^2}}\right\vert_{\{\phi,\delta\} = \{\phi_c,\delta_c\}}\;, \\
N_{3} & \equiv \left.\frac{{{d^2}V_{\rm eff}({\phi },{\delta },{T_{c}})}}{{d\phi d\delta}}\right\vert_{\{\phi,\delta\} = \{\phi_c,\delta_c\}} \;, \\
P_{3} &\equiv \left.\frac{{{d^2}V_{\rm eff}({\phi },{\delta },{T_{c}})}}{{d\delta ^2}}\right\vert_{\{\phi,\delta\} = \{\phi_c,\delta_c\}}  \;.
\end{align}
We estimate the critical temperature and the corresponding classical Higgs field values by requiring
\begin{align}
&V_{\rm eff}(0,0,T_c)  =V_{\rm eff}(\phi_c,\delta_c,T_c) \;,\\
& {\left. {\frac{{d{V_{{\rm{eff}}}}(\phi ,\delta ,{T_c})}}{{d\phi }}} \right|_{\{\phi,\delta\} = \{\phi_c,\delta_c\}}}   = 0, \\
& {\left. {\frac{{d{V_{{\rm{eff}}}}(\phi ,\delta ,{T_c})}}{{d\delta }}} \right|_{\{\phi,\delta\} = \{\phi_c,\delta_c\}}}  = 0 \;.
\end{align}
{\color{black}
In the framework of electroweak baryogenesis, a SFOEWPT is required to ensure the generated baryon number during the phase transition not to be washed out by the electroweak sphaleron process. \yong{Quantitatively, this can be achieved by requiring} $\xi \equiv v/T\geq 1$~\cite{Moore:1998swa,Morrissey:2012db}.\footnote{See Refs.~\cite{Zhou:2019uzq,Zhou:2020xqi} for the condition at the bubble nucleation temperature for different models.} Here, we comment \yong{on} that this condition mostly suffers from
the fluctuation determinant uncertainty \yong{which is comparable to that} in the
lattice simulation of the sphaleron rate\,\cite{Gan:2017mcv,DOnofrio:2014rug}. \yong{For this reason,
in the following sections, we require instead} $\xi\equiv v_c/T_c\geq 1$,
with $v_c$ the critical classical Higgs field values at the critical temperature $T_c$. \yong{For the triplet model}, since $\delta_c \ll \phi_c$ \yong{as discussed above}, we adopt the approximation that $v_c \simeq \phi_c$.}

\subsection{Numerical results}\label{sebsec:nmsetup}

With the discussion presented in last subsection, we then work out the pattern of phase transition in the complex triplet model by scanning over its currently available parameter space. For that purpose, we propose four benchmark setups based on considerations from theoretical constraints on this model discussed in section\,\ref{sec:modelsetup} and the collider results in Ref.\,\cite{Du:2018eaw} for this model:
\begin{itemize}
\item Setup {1}: $\lambda_4\in [-0.5,3], \lambda_5 \in [-3,3], v_\Delta \in [10^{-6}, 10^{-4}]{\rm \,GeV}, M_\Delta \in [0,400]$\,GeV.
\item Setup {2}: $\lambda_4\in [-0.5,3], \lambda_5 \in [-3,3], v_\Delta \in [10^{-6}, 1]{\rm \,GeV}, M_\Delta \in [900,4000]$\,GeV.
\item Setup {3}: $\lambda_4\in [-0.5,3], \lambda_5 \in [-3,3], v_\Delta \in [10^{-5.4}, 1]{\rm \,GeV}, M_\Delta \in [350,900]$\,GeV.
\item Setup {4}: $\lambda_4\in [-0.5,3], \lambda_5 \in [-3,3], v_\Delta \in [10^{-5.4}, 1]{\rm \,GeV}, M_\Delta=500$\,GeV.
\end{itemize}
Furthermore, we fix $\lambda_1=0.129$, $\lambda_2=0.2$ and $\lambda_3=0$ throughout this work. This specific choice of input values for $\lambda_{1,2,3}$ does not lose any generality of our result for the following reason: $\lambda_1$ is basically fixed by the SM Higgs mass, while both $\lambda_2$ and $\lambda_3$ have negligible impact on our conclusion due to the smallness of $v_\Delta$ as discussed above.

\begin{table}[!htp]
\begin{center}
\begin{tabular}{c| c c c c c}
\hline
&~~$\lambda_4$~~& ~~$\lambda_5$~~&~~ $M_\Delta$(GeV)~~&~~$v_\Delta$(GeV)~~&~~$T_c$(GeV) \\
\hline
${\rm BM}_1$ & 1.80& $2.98$ & $379.10$& $5.17\times10^{-6}$&113.81\\
${\rm BM}_2$ & 1.97& 2.29 & $353.51$& $4.63\times10^{-6}$&113.62\\
${\rm BM}_3$ & 2.99& 2.98 & $500.00$& $1.85\times10^{-5}$&145.62\\
\hline
\end{tabular}\caption{Three benchmark points for illustrating the evolution of the effective potential.}\label{tab:data2}
\end{center}
\end{table}

\begin{figure}[!htp]
\begin{center}
\includegraphics[width=0.3\textwidth]{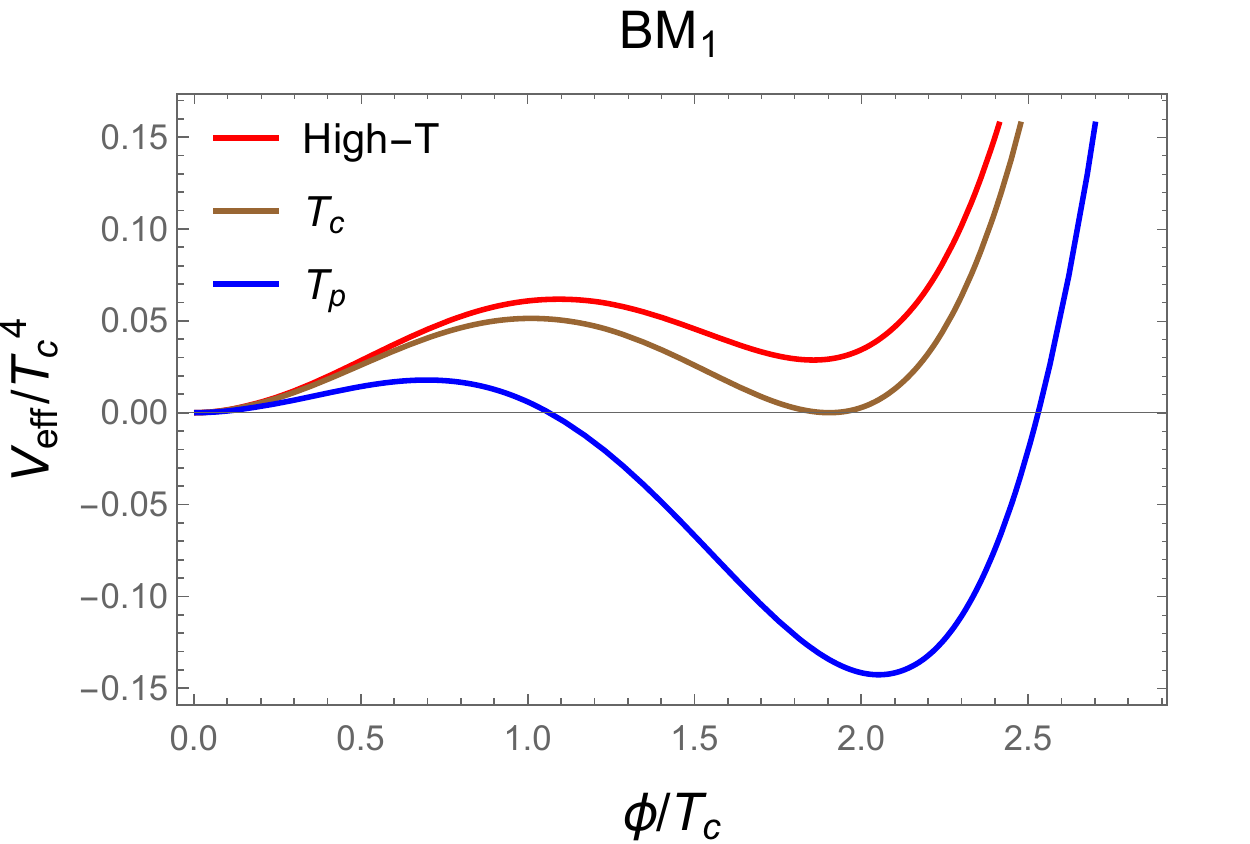}
\includegraphics[width=0.3\textwidth]{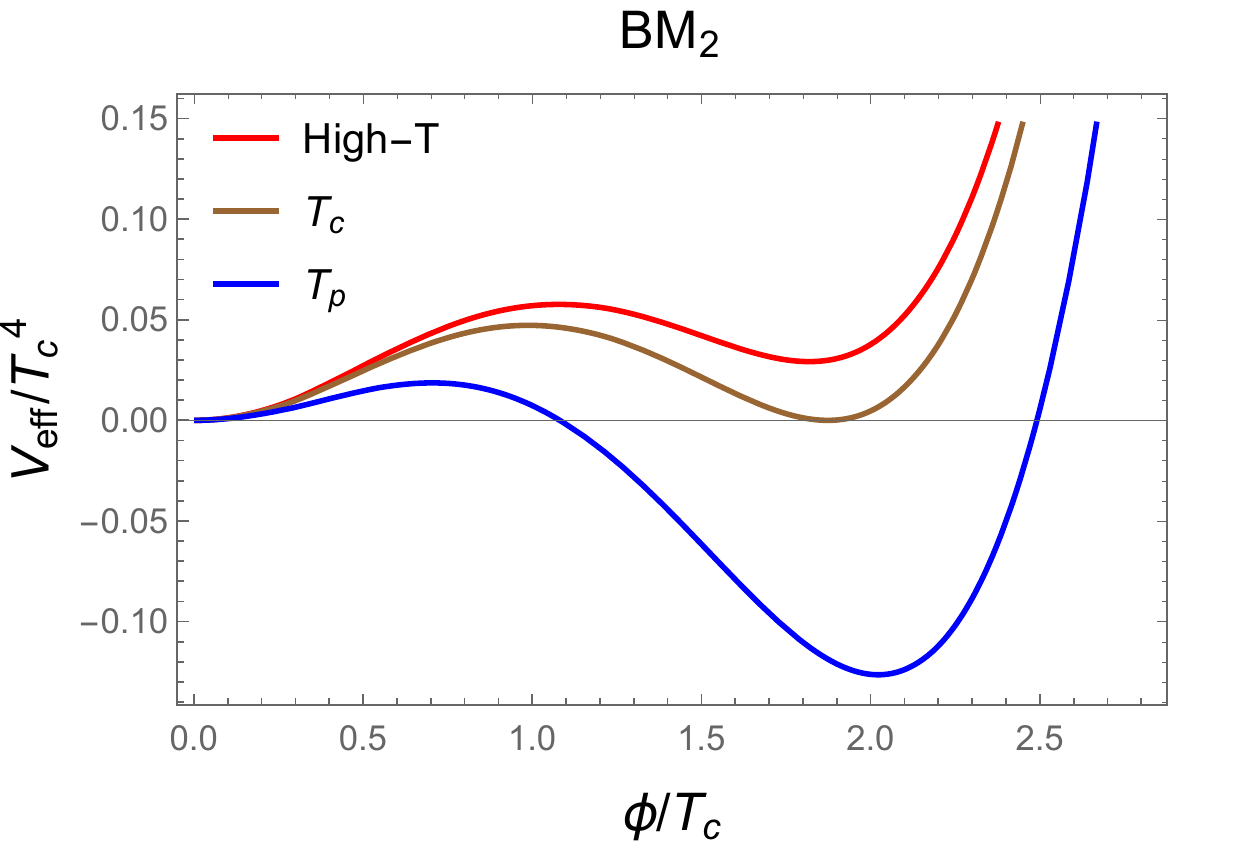}
\includegraphics[width=0.3\textwidth]{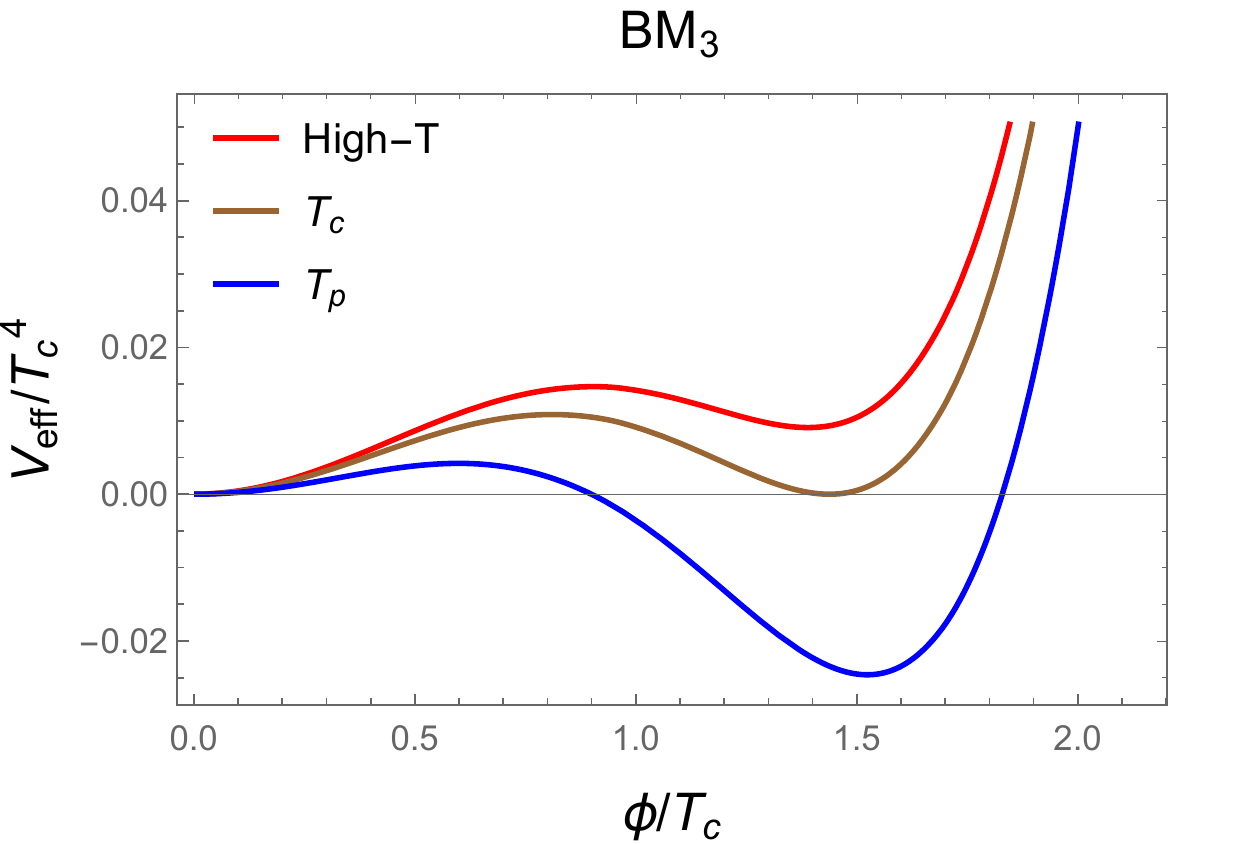}
\includegraphics[width=0.3\textwidth]{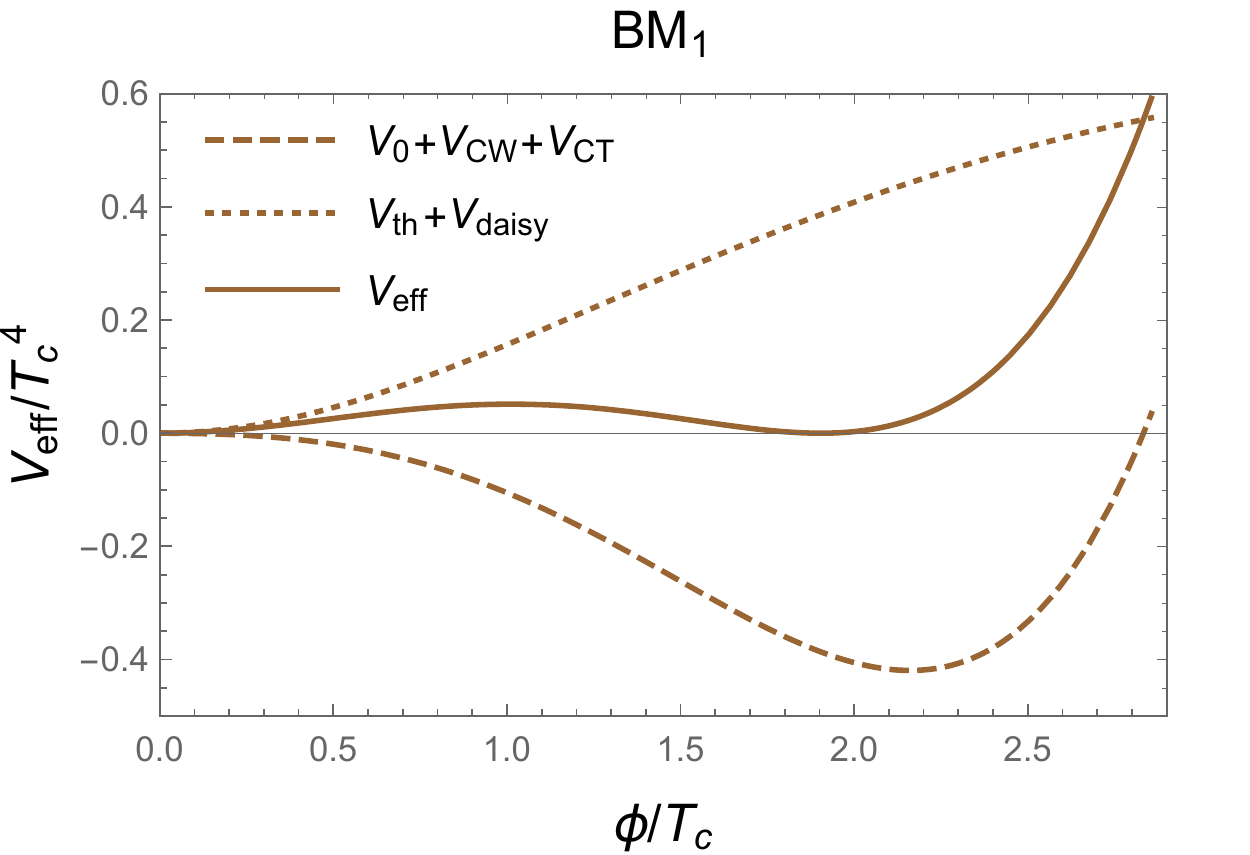}
\includegraphics[width=0.3\textwidth]{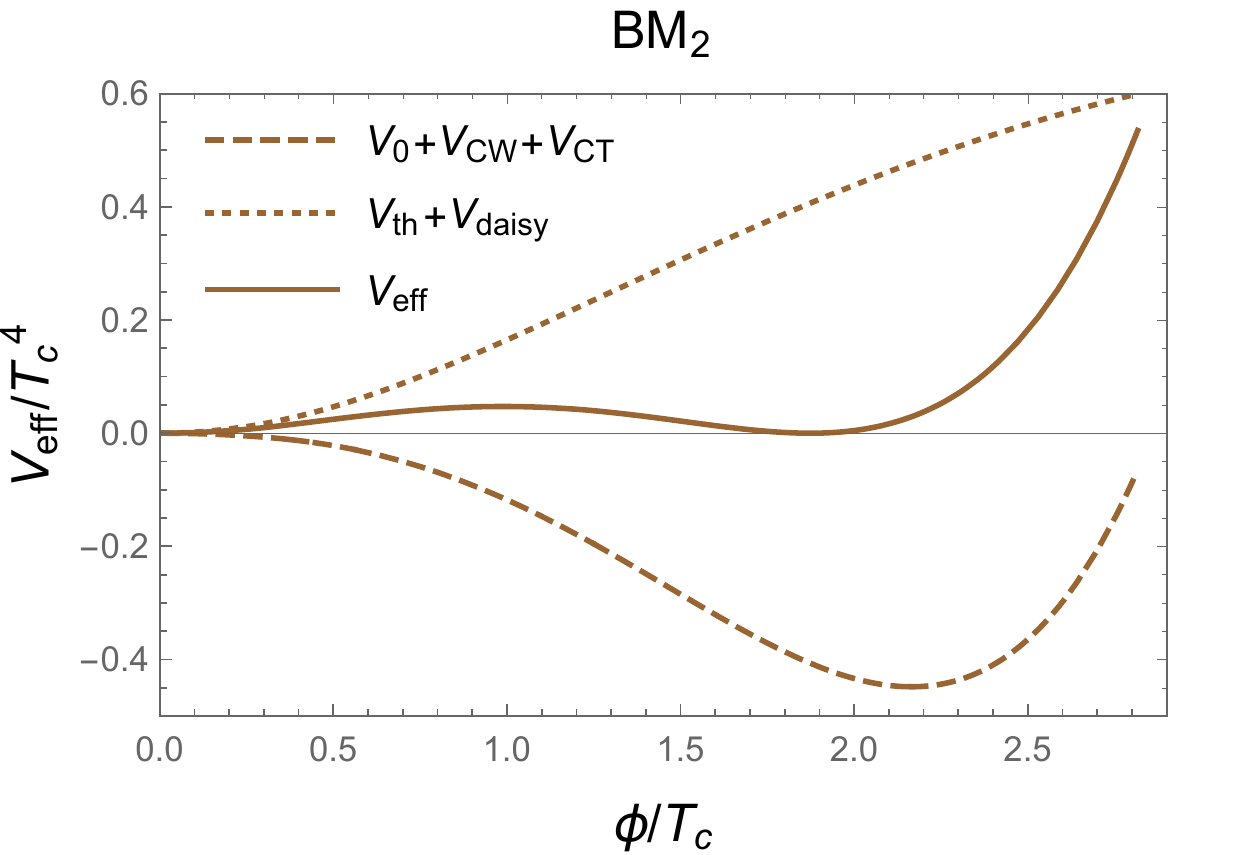}
\includegraphics[width=0.3\textwidth]{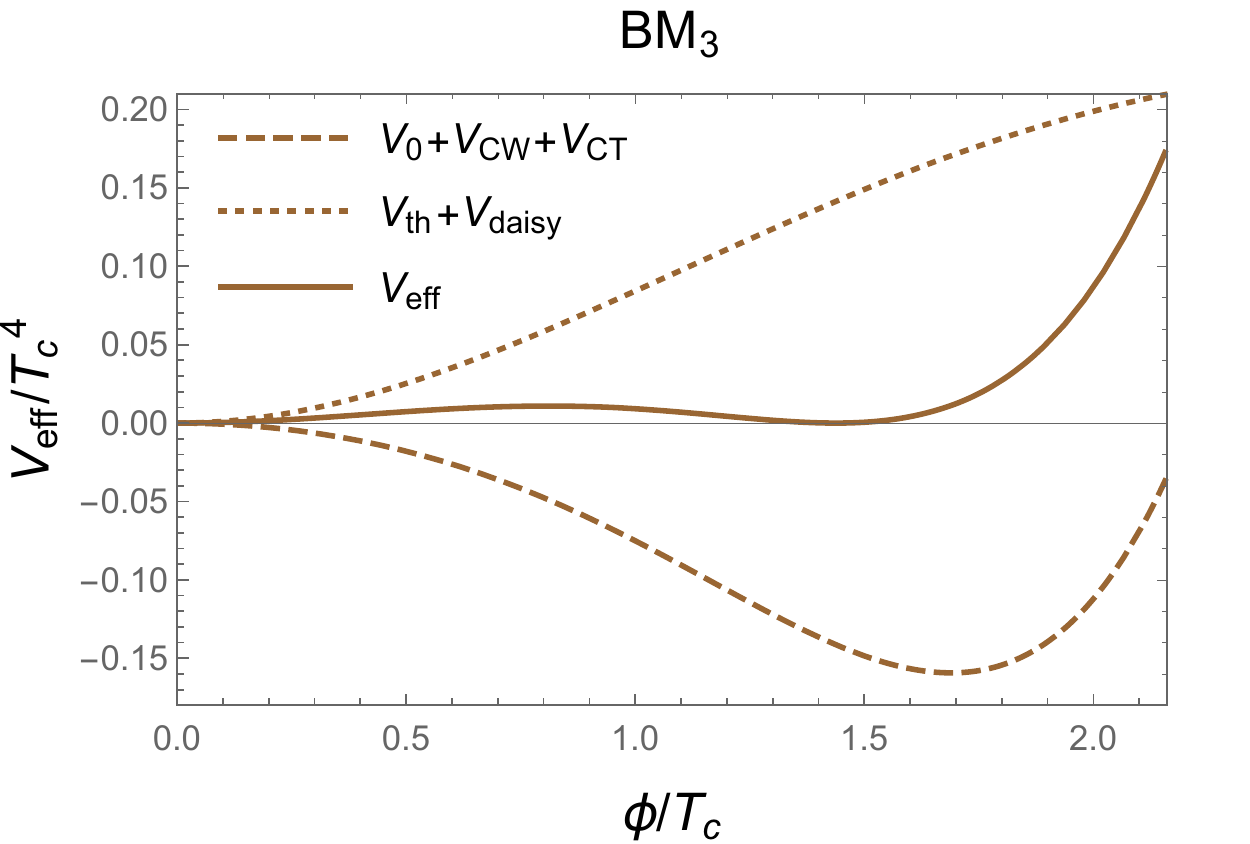}
\end{center}
\caption{Top: Evolution of the vacuum for three benchmark points on the $V_{\rm eff}/T_c^4 -\phi/T_c$ plane. The three temperatures are, respectively, some temperature higher than $T_c$ (red), the critical temperature $T_c$ (brown), and the percolation temperature $T_p$ (blue) as will be further explained in Section~\ref{sec:gravwave}; Bottom: \yong{Components} of the effective potential at $T_c$. }\label{fig:veff}
\end{figure}

{\color{black}  For illustration, we present the thermal effective potential at different temperatures for three benchmark points given in table\,\ref{tab:data2}, \yong{and illustrate in the top row of figure\,\ref{fig:veff} how phase transition occurs}. In each plot, $T_c$ represents the critical temperature, and $T_p$ is the percolation temperature whose definition will become clear in section\,\ref{sec:gravwave}. As \yong{seen from these plots, when }the Universe cools down, the potential barrier \yong{would arise and suggest} that the phase transition is of first order. \yong{Accordingly, we show in the bottom row of figure\,\ref{fig:veff} the components of $V_{\rm eff}$ to clarify the fact that the barrier} indeed comes from thermal corrections. \yong{Therefore,} this kind of phase transition \yong{would} belong to the thermally driven class of the electroweak phase transition~\cite{Chung:2012vg}. \yong{Moreover,} our result shows that $V_{\rm th}$ ($V_{\rm daisy}$) contributes positively (negatively) to $V_{\rm eff}$ around the true vacua, \yong{while} $V_{\rm th}+V_{\rm daisy}$ \yong{contributes a net positive correction to} $V_{\rm eff}$. \yong{As a result, }these two thermal corrections lift up the zero-temperature effective potential ($V_0+V_{\rm CW}+V_{CT}$) to \yong{an extent that helps} form a maximum in the potential shape and yields the potential barrier around $\phi/T_c\sim 1$.
}

\begin{figure}[!htp]
\centering{
\begin{adjustbox}{max width = \textwidth}
\begin{tabular}{cc}
\includegraphics[width=0.4\textwidth]{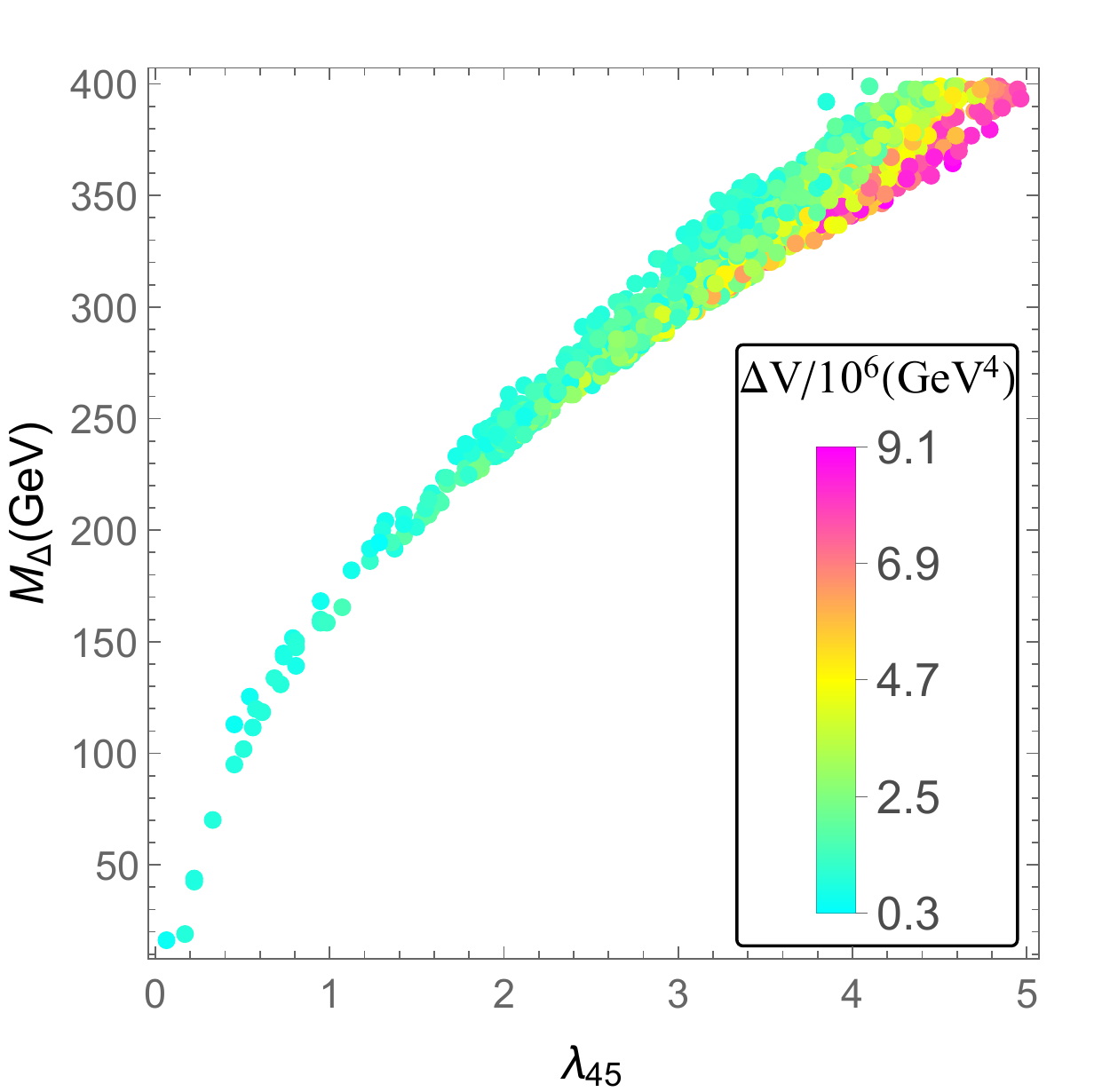} &
\includegraphics[width=0.4\textwidth]{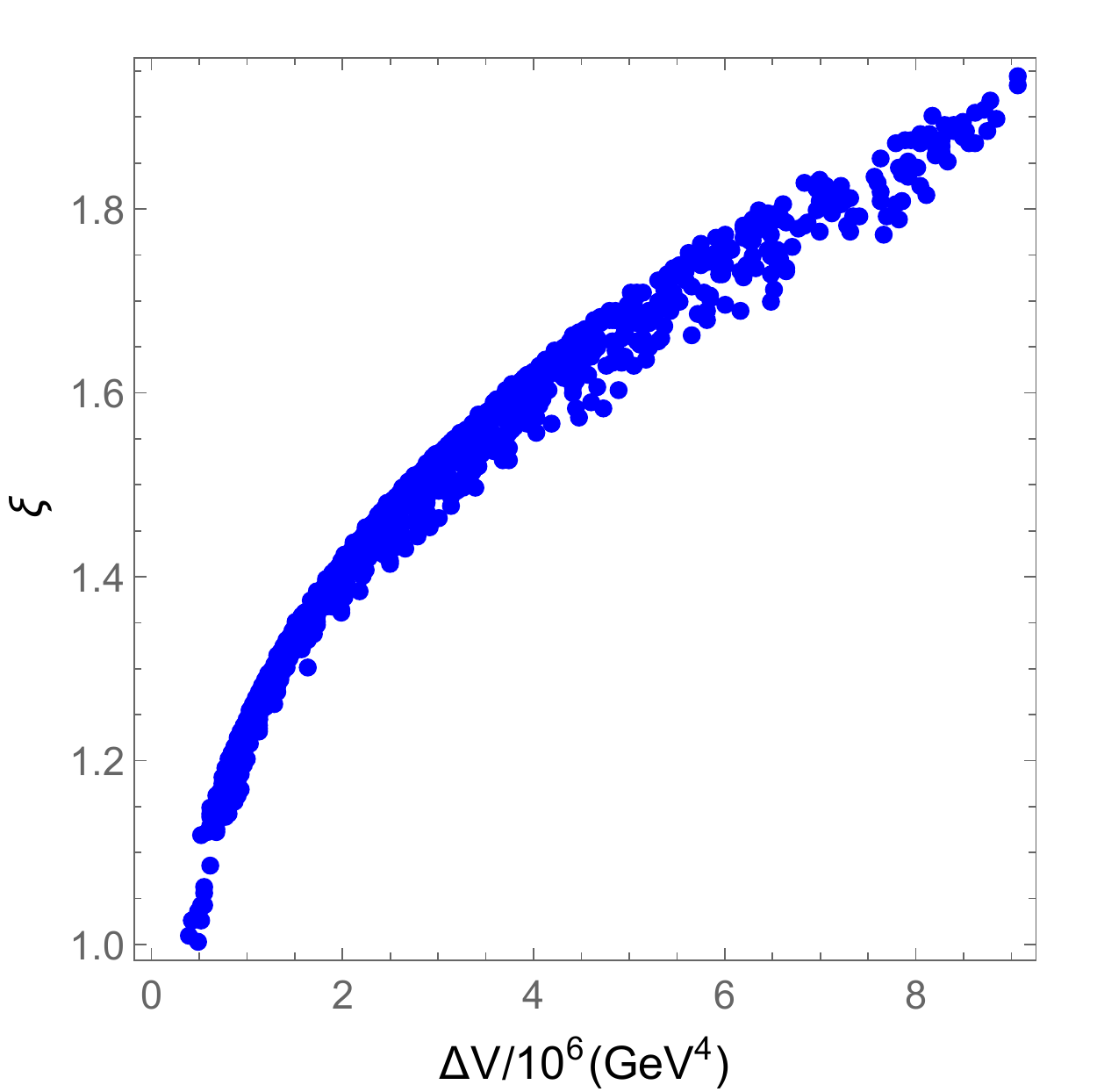}\\
\hspace{4cm}(Setup 1)\hspace{-3cm} \\
\includegraphics[width=0.4\textwidth]{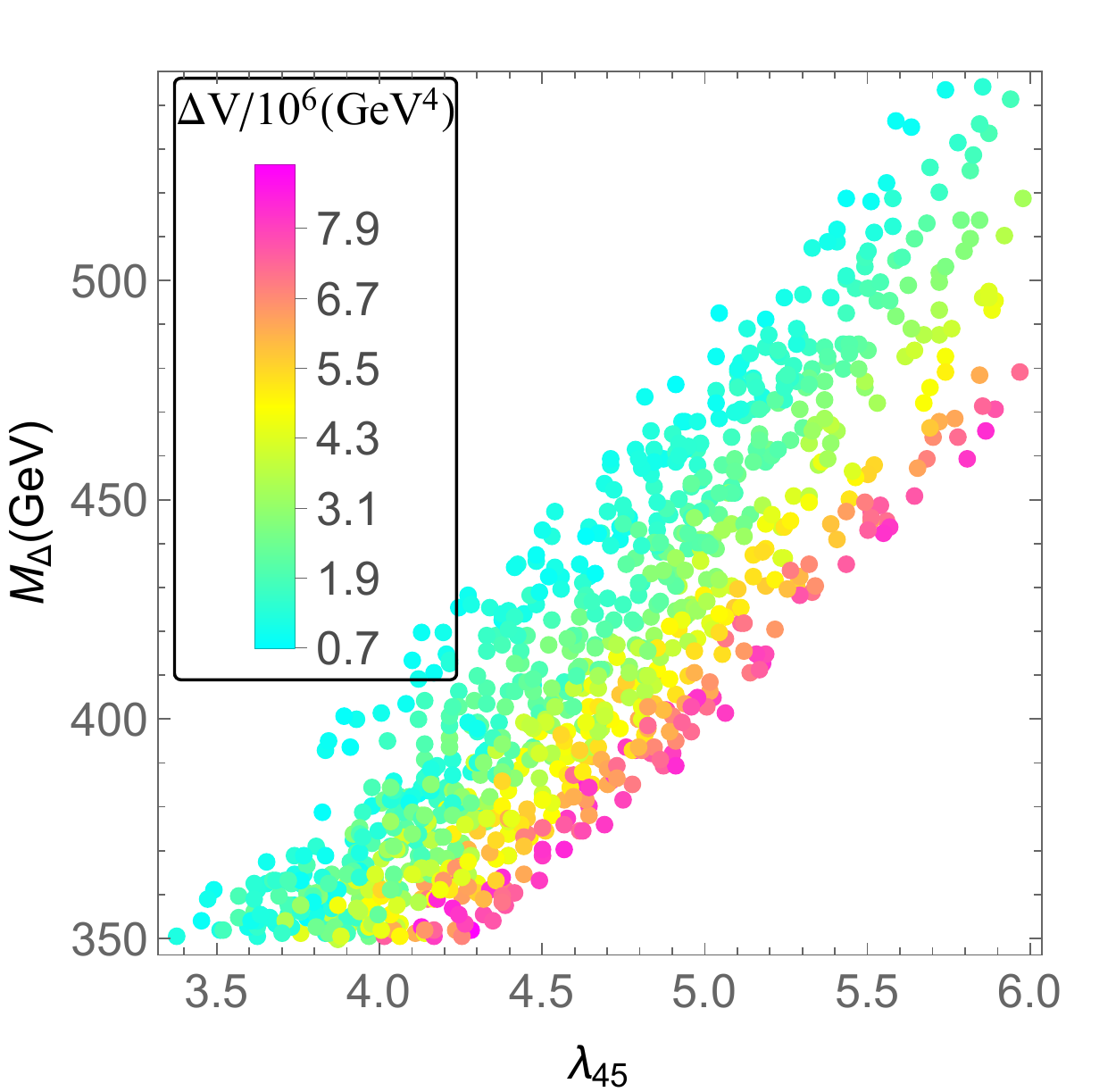} &
\includegraphics[width=0.4\textwidth]{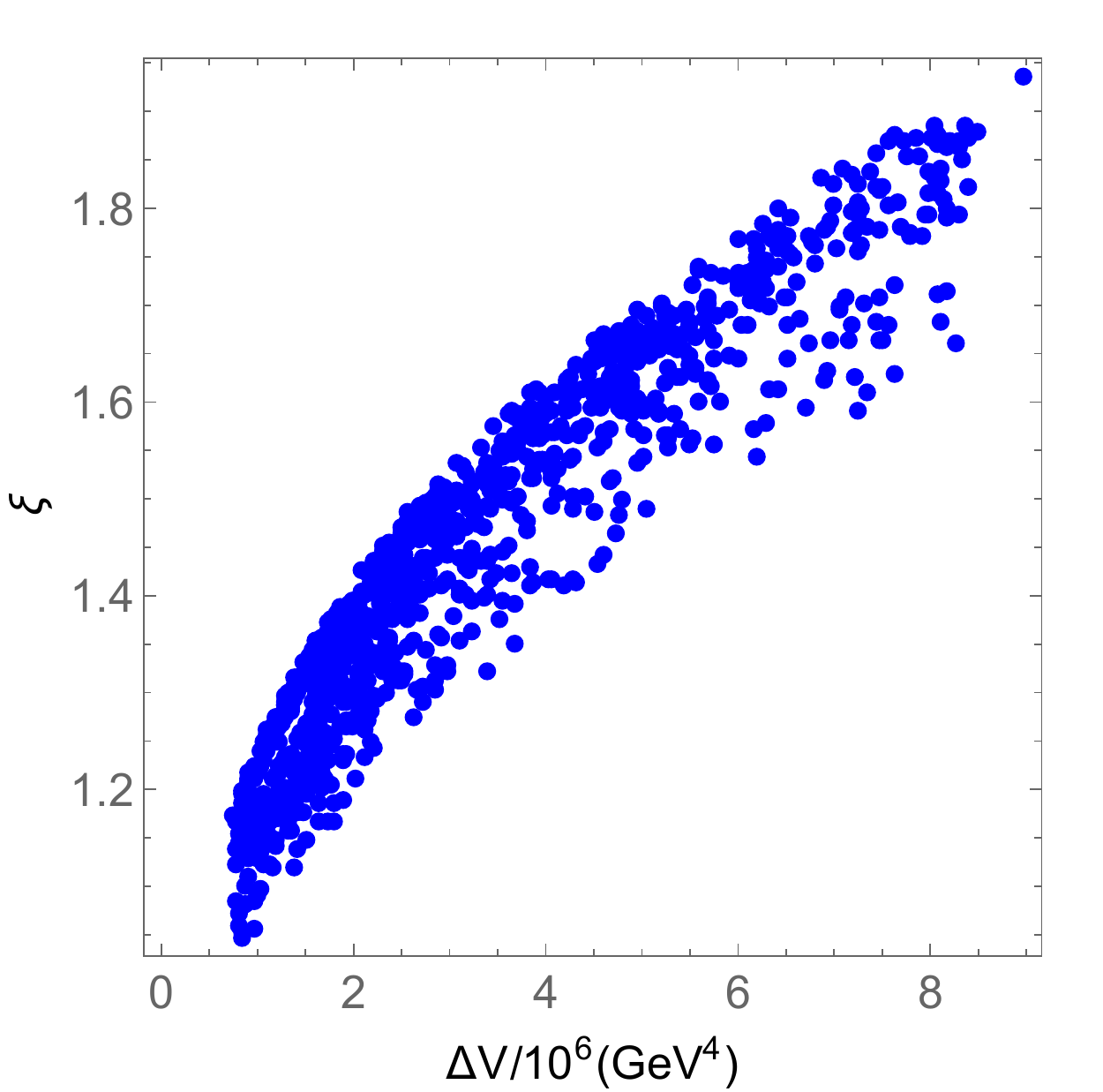}\\
\hspace{4cm}(Setup 3)\hspace{-3cm} \\
\includegraphics[width=0.4\textwidth]{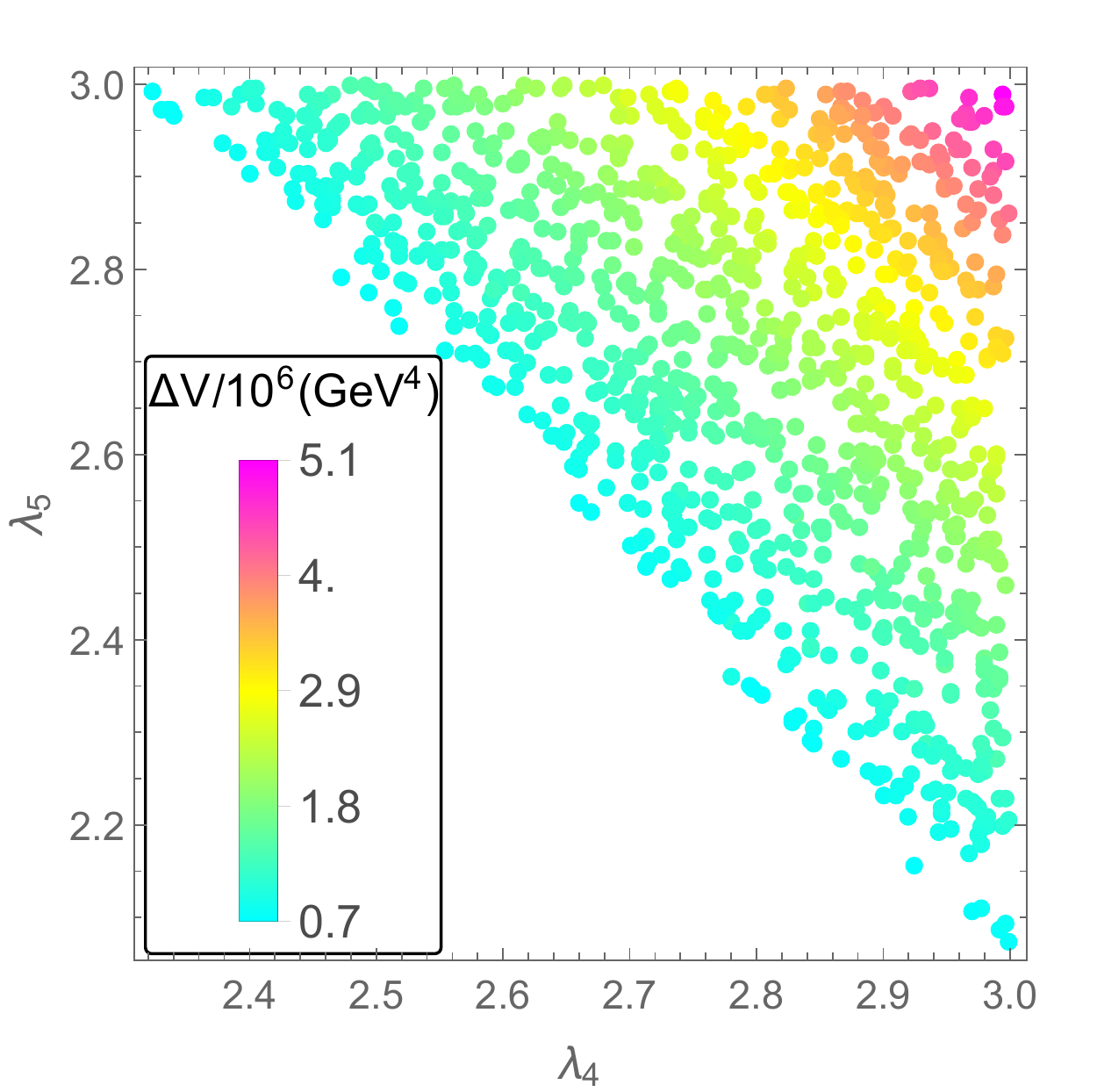} &
\includegraphics[width=0.4\textwidth]{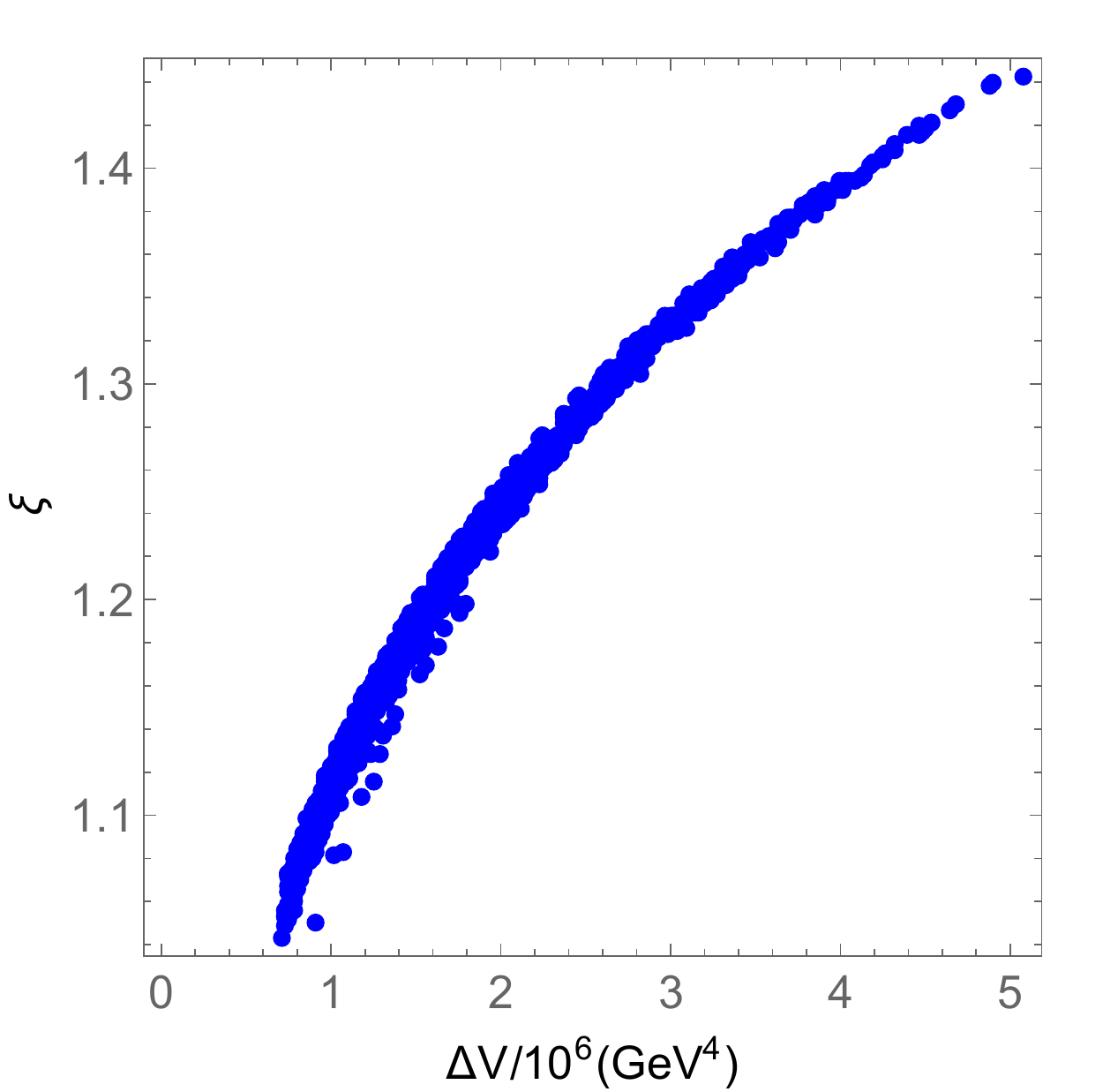}\\
\hspace{4cm}(Setup 4)\hspace{-3cm}
\end{tabular}
\end{adjustbox}
}
\caption{The height of the potential barriers and the $\xi$ for various $\lambda_{45}$ and $M_\Delta$ for the four setups. Note that setup 2 is missing due to the fact that the decoupling of the heavy triplet.}\label{fig:sfoewpt}
\end{figure}

To ensure the first-order phase transition is strong enough to avoid later time washout of the baryon numbers, one needs $\xi>1$ as discussed above. For a successful SFOEWPT in the type-II model, after performing the numerical calculations, we show our results in figure\,\ref{fig:sfoewpt} for the four setups above, where the left column shows the results for the barrier height, and the right column for $\xi$. Note that the second setup is missing in our results due to the fact that the triplet scalars in this scenario are too heavy to contribute to the potential barrier and therefore decouple from the phase transition. For setup 1 and 3, the height of the barriers are only functions of $\lambda_{45}$ and $M_\Delta$. While for setup 4, since we fix $M_\Delta=500$\,GeV, we plot the barrier height as a function of the individual $\lambda_{4,5}$ couplings instead. Clearly, our results show that large $\lambda_{4,5}$ and heavy $M_\Delta$ help {\yong{increase}} the barrier height and therefore enhance \yong{the}
value of $\xi$ as seen from the second column of figure\,\ref{fig:sfoewpt}. However, we comment on that due to the decoupling effects, when $M_\Delta$ exceeds $\sim550$\,GeV, the barrier height would become insufficient to induce a SFOEWPT as implied in the first two plots in the left column of figure\,\ref{fig:sfoewpt}. Interestingly, most of the $\xi \geq 1 $ viable points falls into the mass region that could be tested at current/future colliders\,\cite{Du:2018eaw}, suggesting the possible synergy of different probes in searching for the type-II seesaw model. For each setup, we further discuss this possibility below.

\begin{figure}[!htp]
\centering{
\begin{adjustbox}{max width = \textwidth}
\begin{tabular}{cc}
\includegraphics[width=0.4\textwidth]{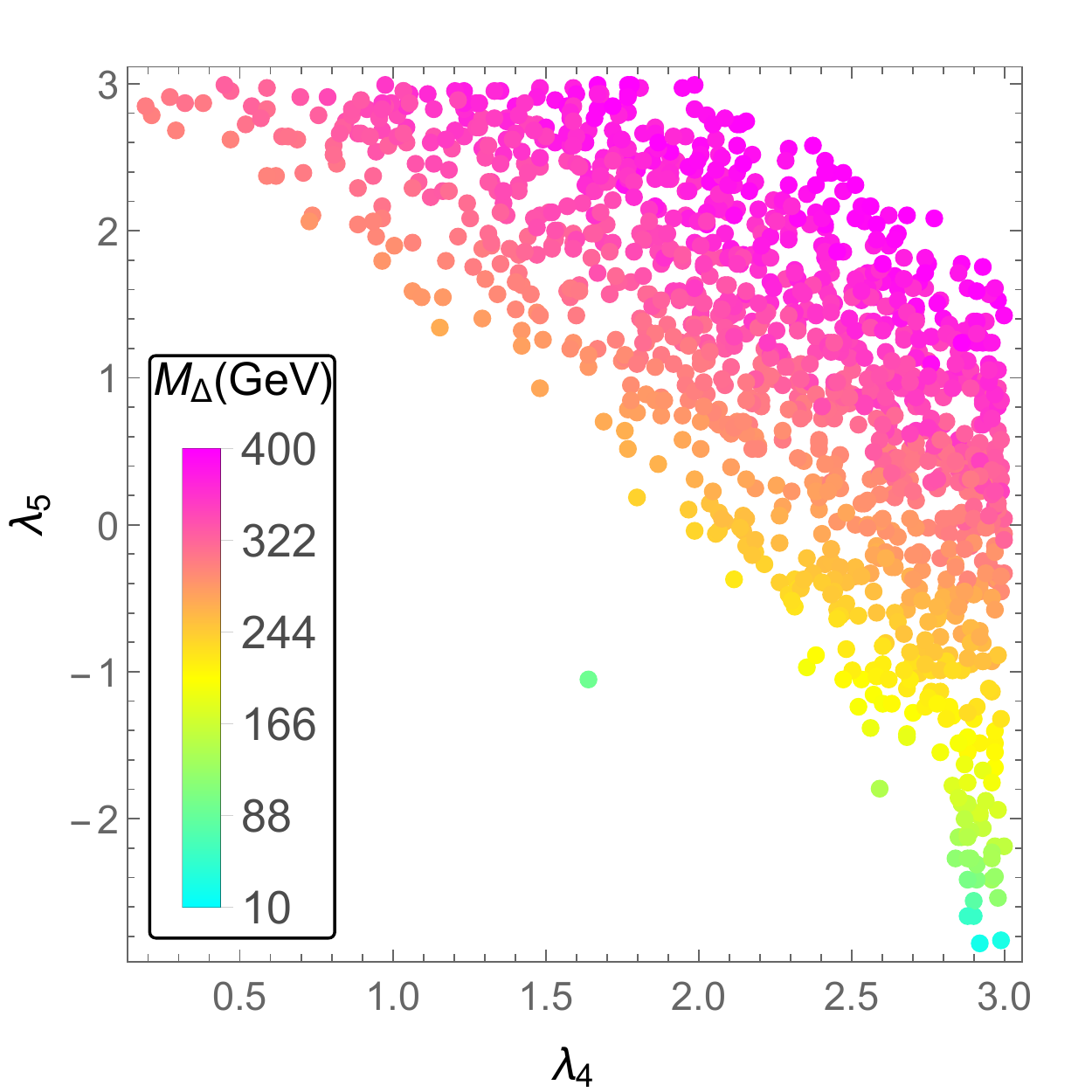} &
\includegraphics[width=0.4\textwidth]{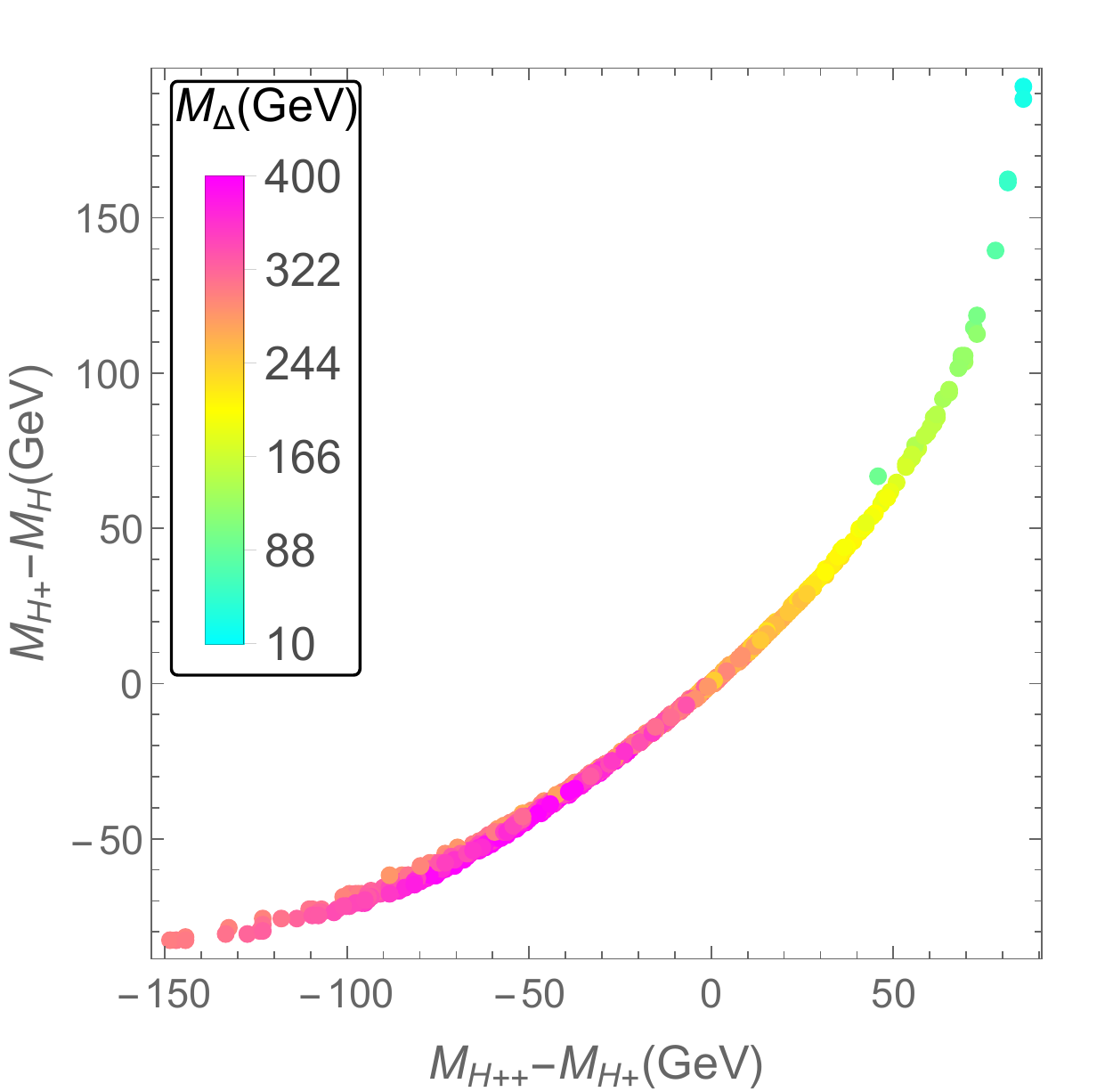} \\
\includegraphics[width=0.4\textwidth]{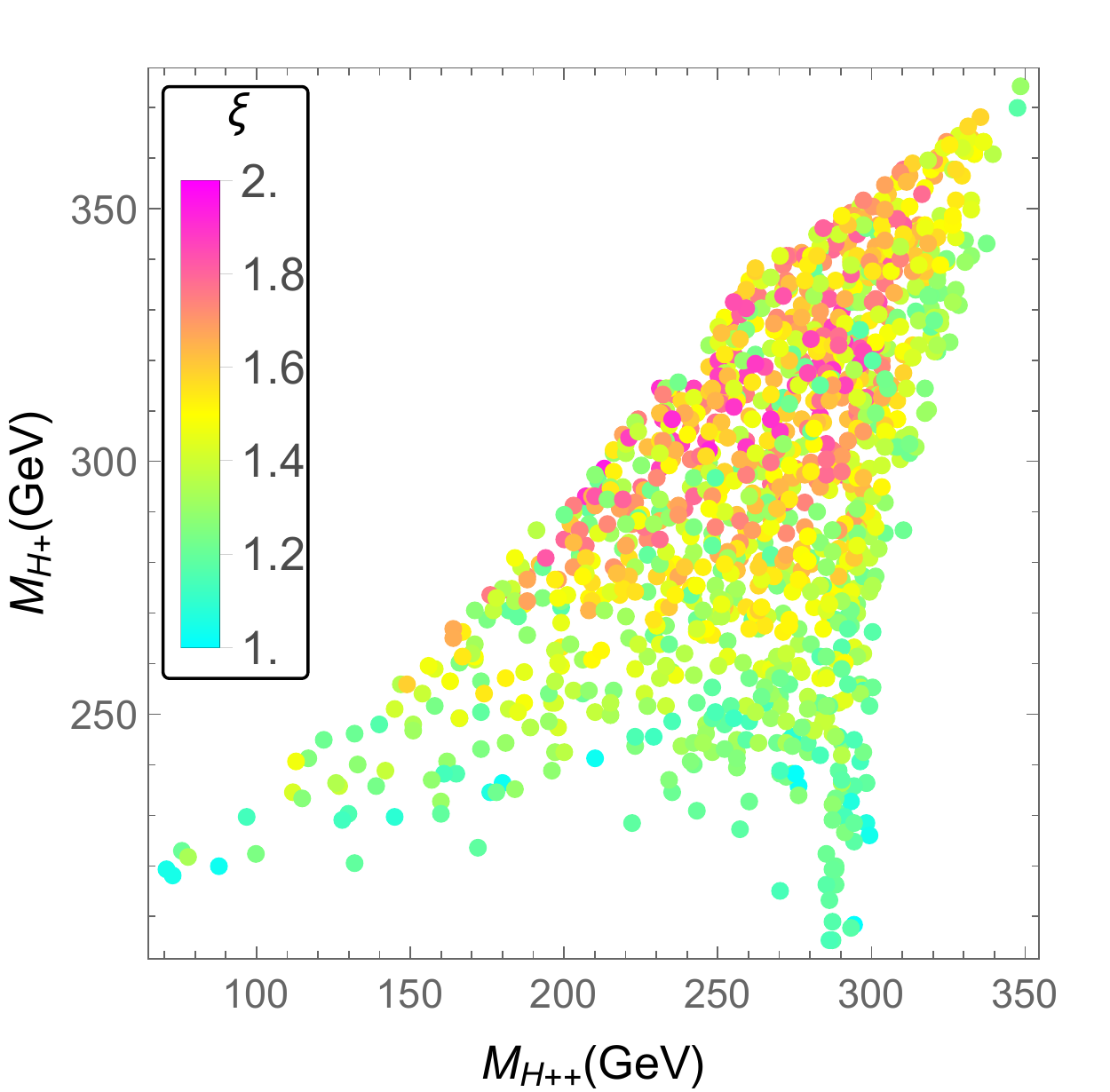}  &
\includegraphics[width=0.4\textwidth]{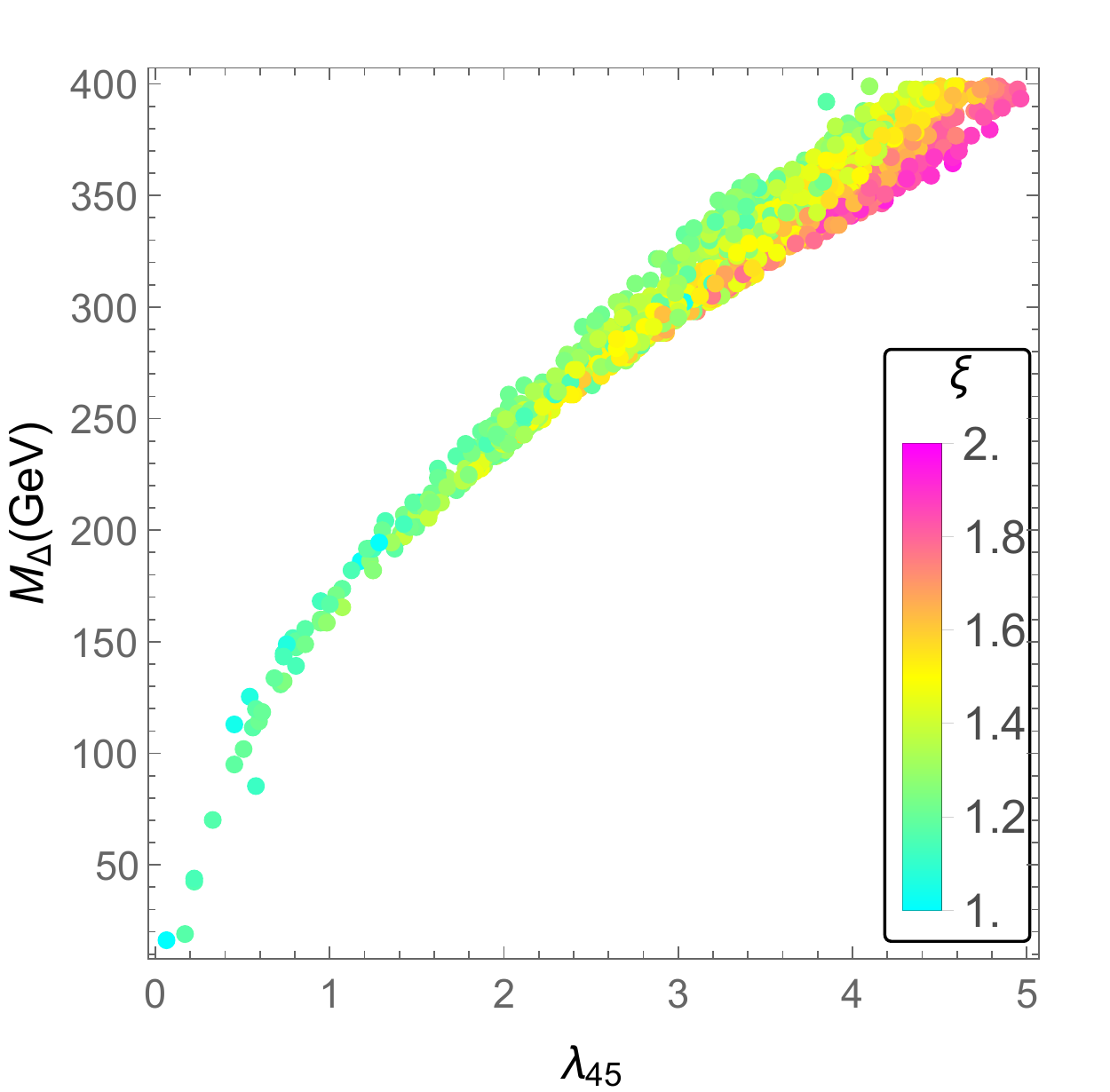}
\end{tabular}
\end{adjustbox}
}
\caption{SFOEWPT viable benchmark points for our setup 1. In the upper row, we show benchmark points with $\xi>1$, while in the second row, we show explicitly the value of $\xi$ on different parameter planes. See the main text for details.}\label{fig_setup1Q174}
\end{figure}

For setup 1, our results are shown in figure\,\ref{fig_setup1Q174}. The first plot in the upper row shows the benchmark points for a SFOEWPT with varying $M_\Delta$ and $\lambda_{4,5}$. Note that a light triplet with $M_\Delta\simeq10$\,GeV still permits a SFOEWPT and we comment on that the triplet mass eigenvalues are much larger than $M_\Delta$ in this case due to corrections from negative $\lambda_5$'s. See our eqs.\,\eqref{mhpp} and \eqref{mhp}, for example. However, we point out that since $v_\Delta$ is small in this case, the same-sign dilepton channel dominates the decay of $H^{\pm\pm}$ and one can thus utilize the $pp\rightarrow H^{++}H^{--}\rightarrow\ell^+\ell^+\ell'^-\ell'^-$ channel to constrain the light triplet scenario. See, for example, Ref.\,\cite{Du:2018eaw}. Similarly, we show in the second plot in the first row of figure\,\ref{fig_setup1Q174} for the benchmark points that can result in a SFOEWPT with different triplet mass differences. The mass differences are essentially only dependent on $\lambda_5$ since $v_\Delta\ll v_\Phi$ and $\sin\alpha\approx0$. In this case, we find the parameter benchmark points are rather limited, suggesting the fact that one could possibly recast current/future experimental results onto the mass difference plane as we show here to determine the mass scale of the triplet and also $\lambda_5$. This in turn would help {\yong{identify}} the triplet model and its model parameter determination at colliders.

On the other hand, the benchmark points for a SFOEWPT are shown in the bottom row of figure\,\ref{fig_setup1Q174}, where the colored legend indicates directly the \yong{value of} $\xi$ \yong{for} the phase transition. Note that the points mainly reside in the lower half of the $M_{H^\pm}-M_{H^{\pm\pm}}$ plane as seen from the first plot of the bottom row. In particular, $\xi$ approaches larger values when $200\lesssim M_{H^\pm}\lesssim300$\,GeV and $250\lesssim M_{H^{\pm\pm}}\lesssim350$\,GeV, indicating that positive $\lambda_5$'s are slightly favored for a SFOEWPT as is also clear in the {\color{black}first} plot of the first row. Similarly, from the last plot in the last row of figure\,\ref{fig_setup1Q174}, one sees that positive $\lambda_{45}$ are preferred for a SFOEWPT, suggesting also a preference of positive $\lambda_4$'s that all together help stabilize the Higgs potential up to the Planck scale as observed in Ref.\,\cite{Du:2022vso}.

\begin{figure}[!htp]
\centering{
\begin{adjustbox}{max width = \textwidth}
\begin{tabular}{cc}
\includegraphics[width=0.4\textwidth]{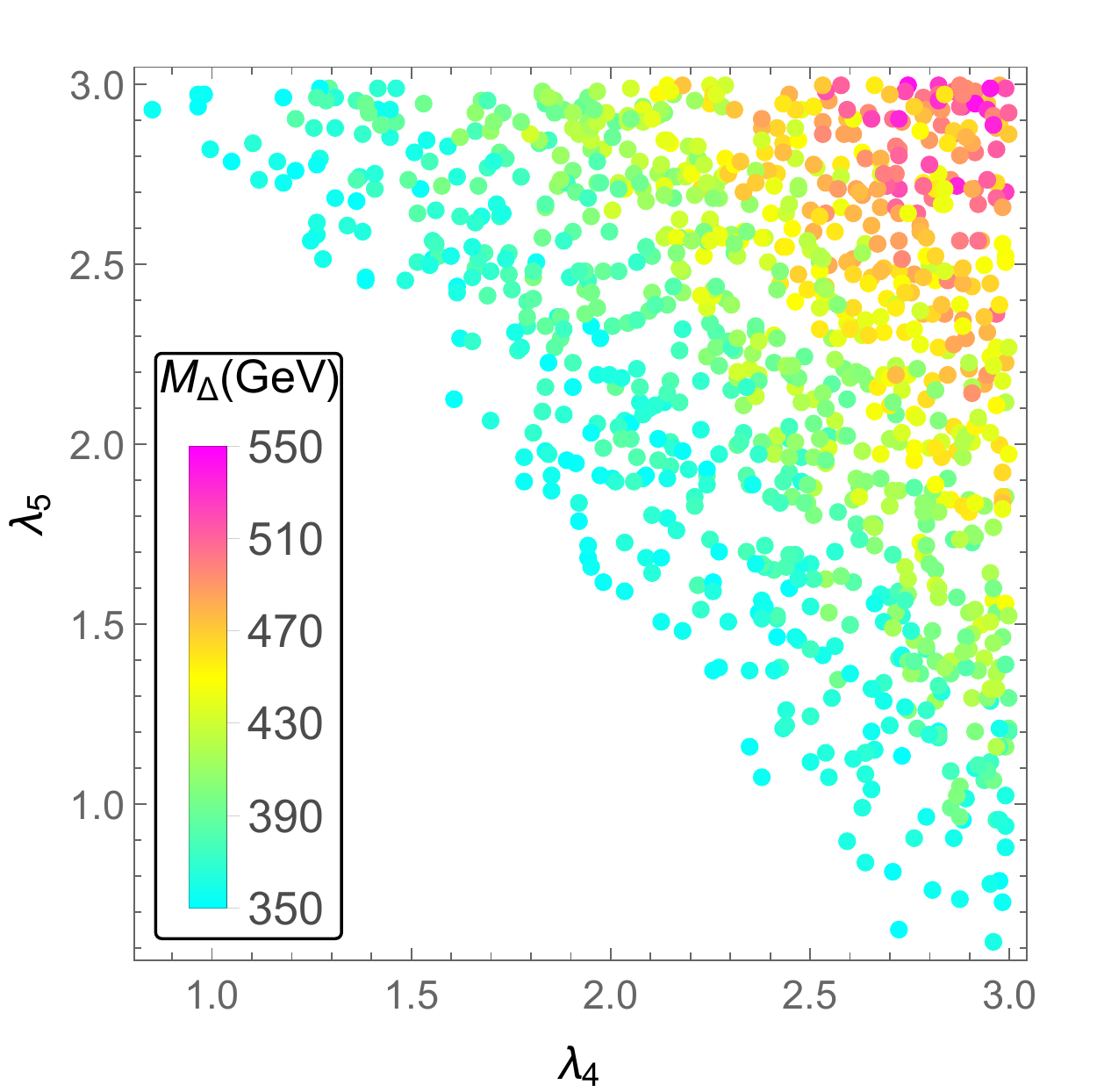} &
\includegraphics[width=0.4\textwidth]{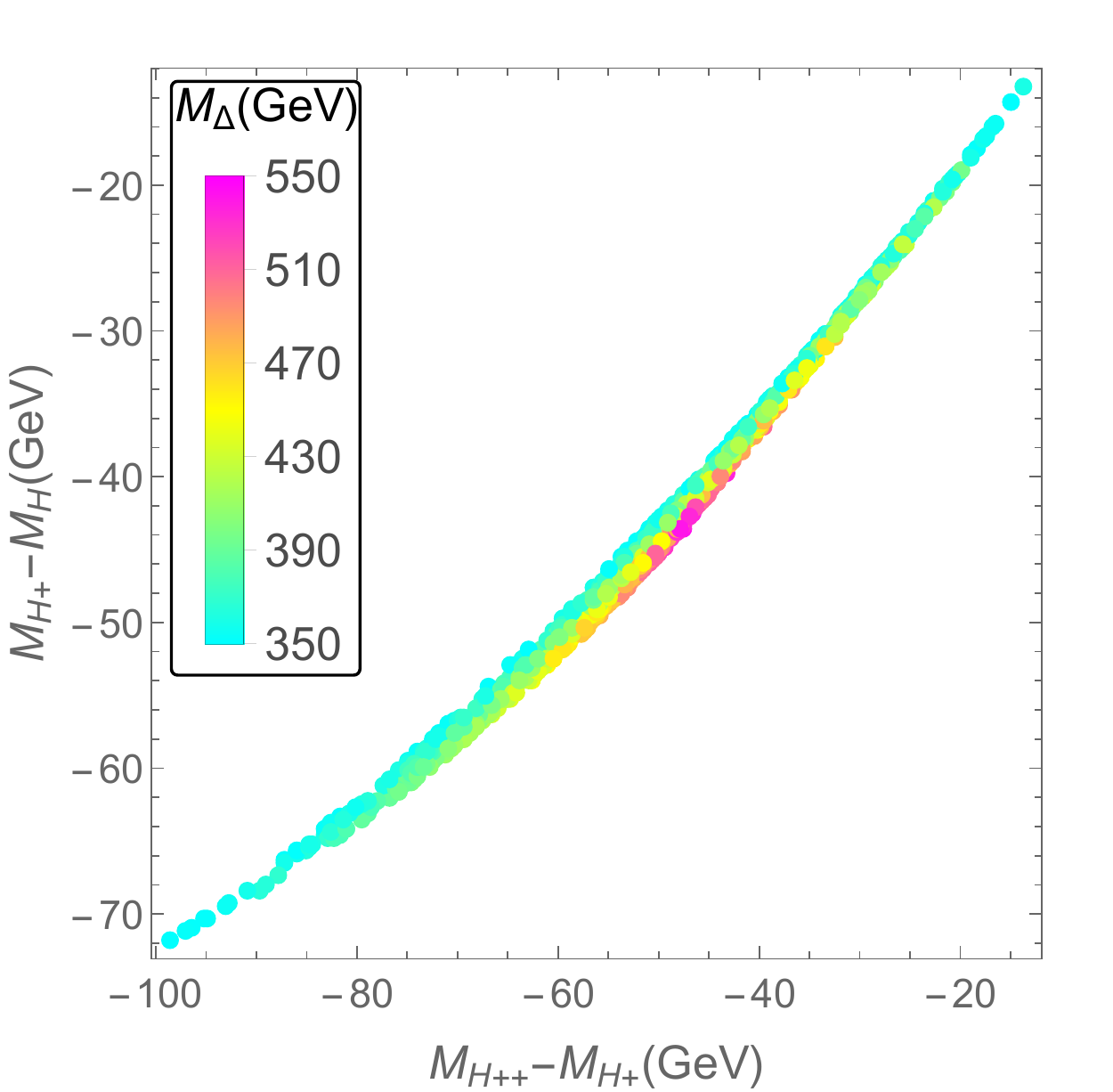}\\
\includegraphics[width=0.4\textwidth]{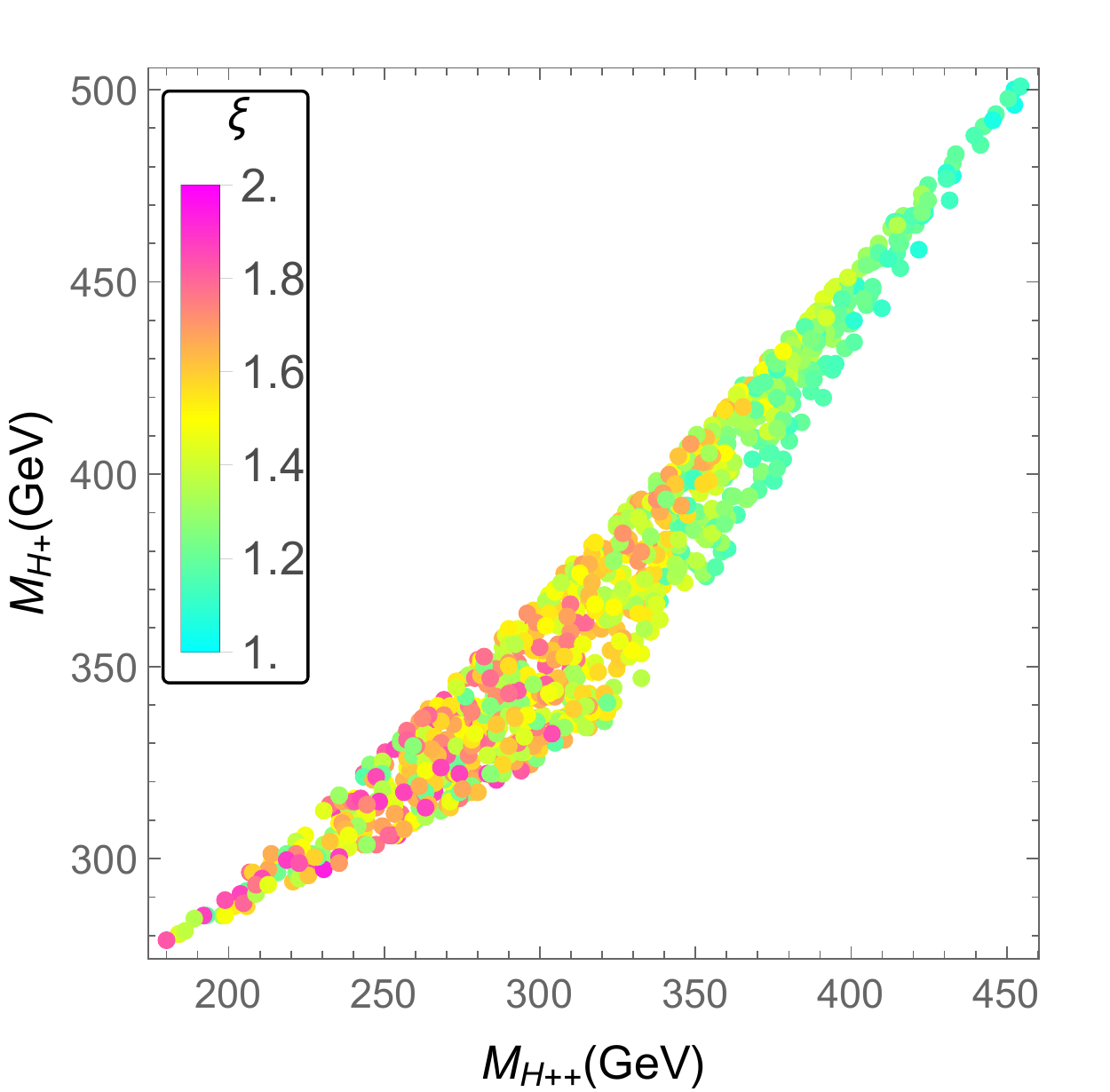} &
\includegraphics[width=0.4\textwidth]{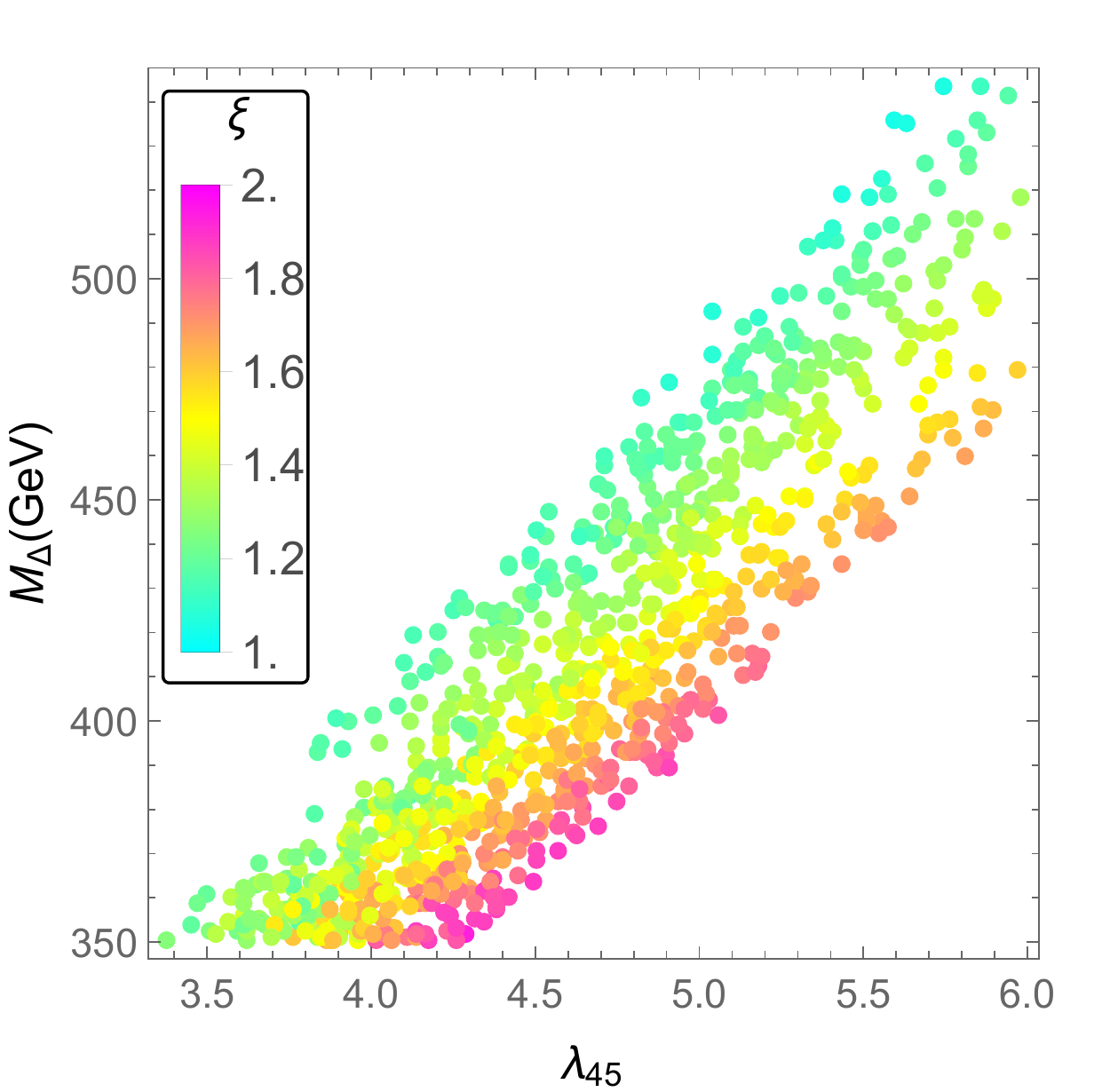}
\end{tabular}
\end{adjustbox}
}
\caption{Same as figure\,\ref{fig_setup1Q174} but for our setup 3.}\label{fig_setup3Q174}
\end{figure}

A similar observation as discussed above applies to our setup 3, which can be seen directly from our figure\,~\ref{fig_setup3Q174}. Note that in our setup 3, even though we scan over a relatively large range of $M_\Delta$ up to about 1\,TeV, light triplet Higgs particles are generically preferred for a SFOEWPT as indicated by the dots in red/purple.

\begin{figure}[!htp]
\centering{
\begin{adjustbox}{max width = \textwidth}
\begin{tabular}{cc}
\includegraphics[width=0.4\textwidth]{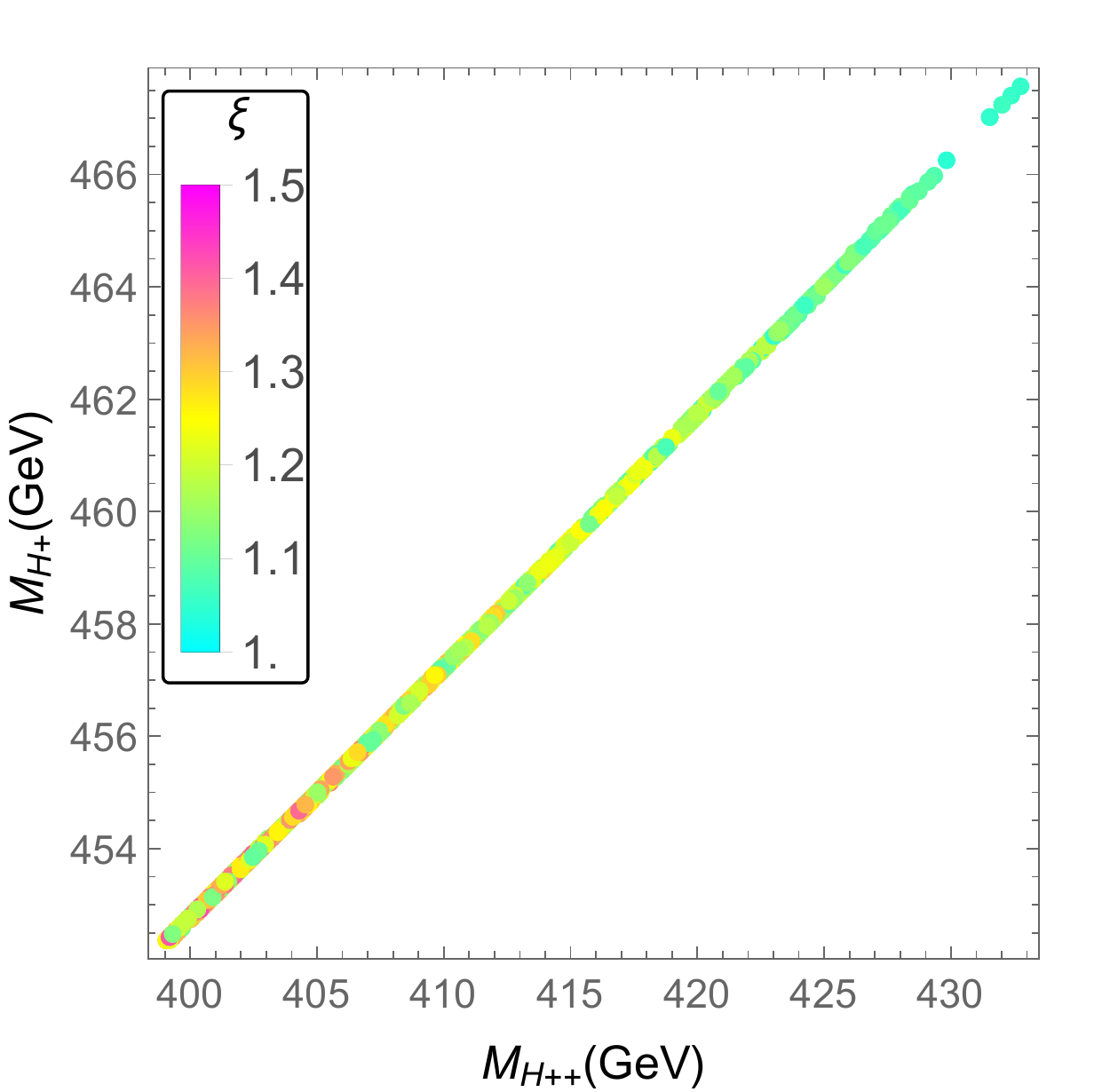} &
\includegraphics[width=0.4\textwidth]{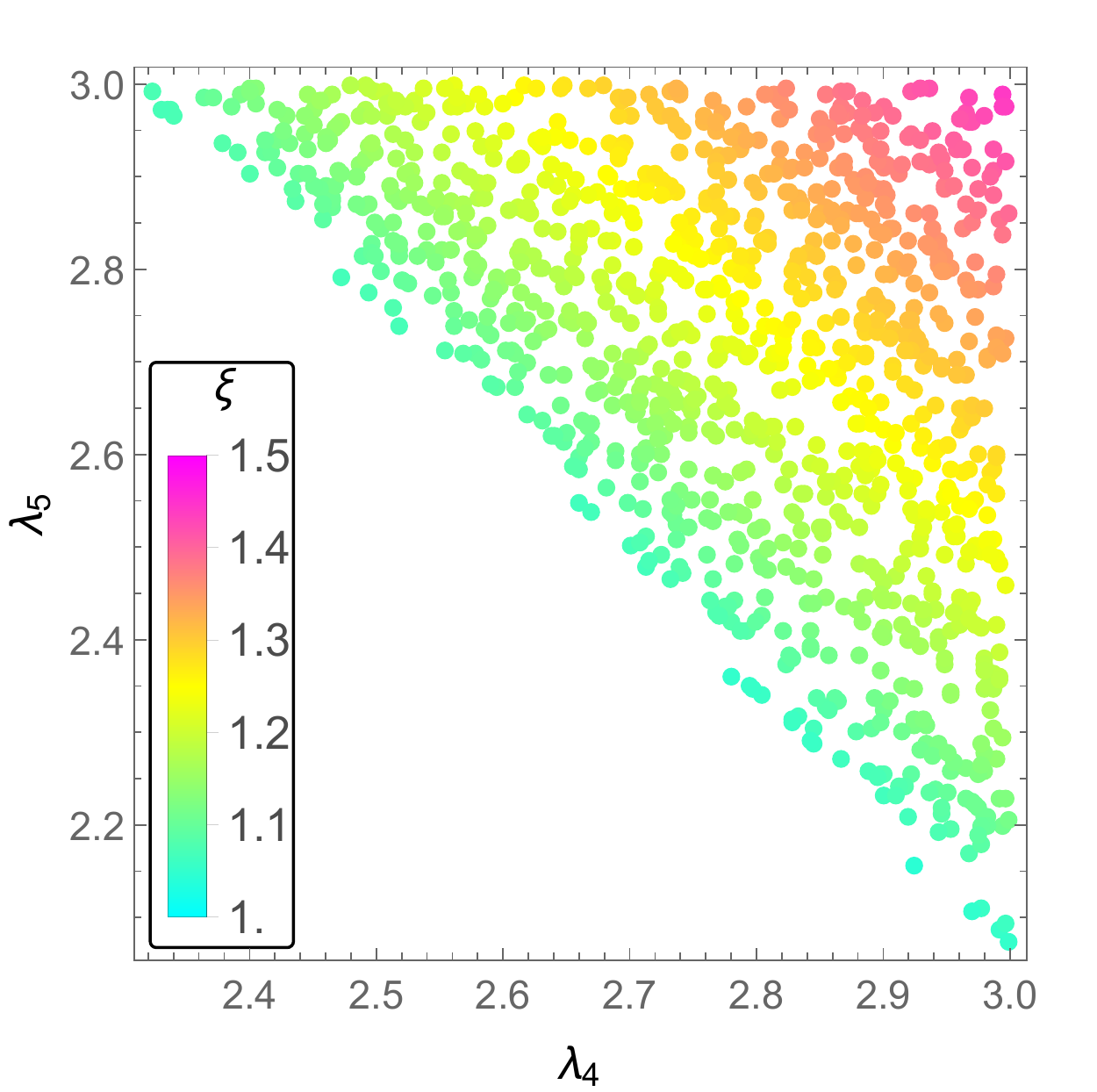}
\end{tabular}
\end{adjustbox}
}
\caption{Left panel: same as the bottom-left of figure\,\ref{fig_setup1Q174} but for our setup 4; Right panel: the explicitly value of $\xi$ on the plane of $\lambda_{4}$ and $\lambda_5$.}\label{fig_setup4Q174}
\end{figure}

Finally, for our setup 4, the results are presented in figure\,\ref{fig_setup4Q174}. Note that in this case, we fix $M_\Delta=500$\,GeV. This is motivated by the consideration that, upon model discovery, around this specific value for example, one can then readily recast the masses of the triplet Higgs particles onto the first panel of figure\,\ref{fig_setup4Q174} to check the existence of a SFOEWPT. From the distribution of $\xi$, one can then utilize the second plot of figure\,\ref{fig_setup4Q174} to possibly determine the sign of $\lambda_{4,5}$, and thus the mass spectrum of the triplet model. We comment on that for setup 4, we again find that positive $\lambda_{4,5}$ are preferred for a SFOEWPT.\footnote{Positive $\lambda_5$'s would correspond to the reversed mass hierarchy discussed in\,\cite{Du:2018eaw}, which can be investigated through the multilepton signatures at hadron colliders\,\cite{Mitra:2016wpr}.}

\subsection{Implications from Br($h\to\gamma\gamma$)}
As already observed in Ref.\,\cite{Chao:2012mx,Du:2018eaw}, theoretical constraints on the portal couplings $\lambda_{4,5}$ are already very stringent, especially for those from one-loop perturbativity summarized in section\,\ref{subsubsec:thcons}. For this reason, we ask ourselves the following question: How could these points obtained in last subsection that are responsible for a SFOEWPT could be tested from current and/or future collider experiments?

\begin{figure}[!htp]
\begin{center}
\includegraphics[width=0.4\textwidth]{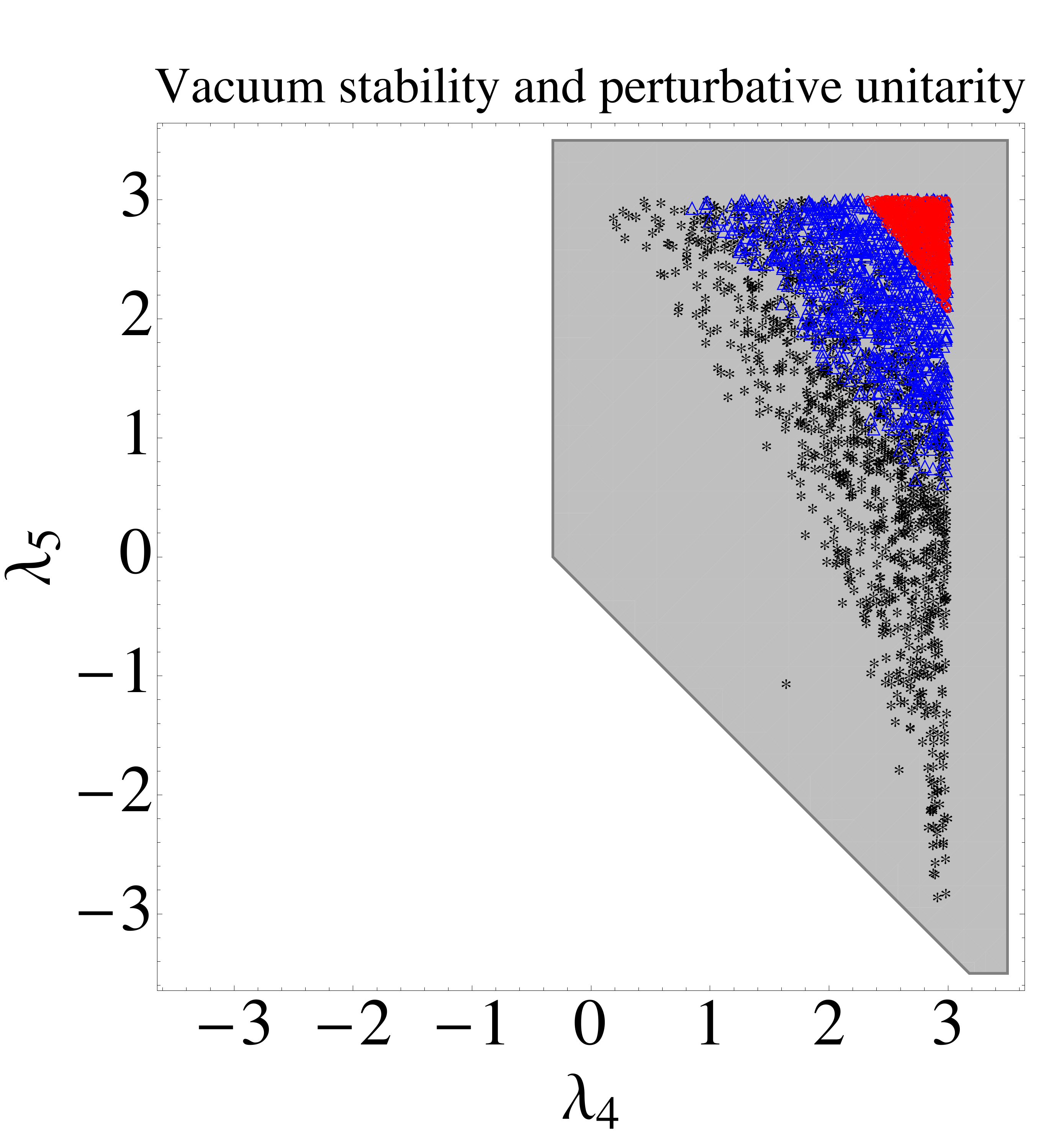} \includegraphics[width=0.45\textwidth]{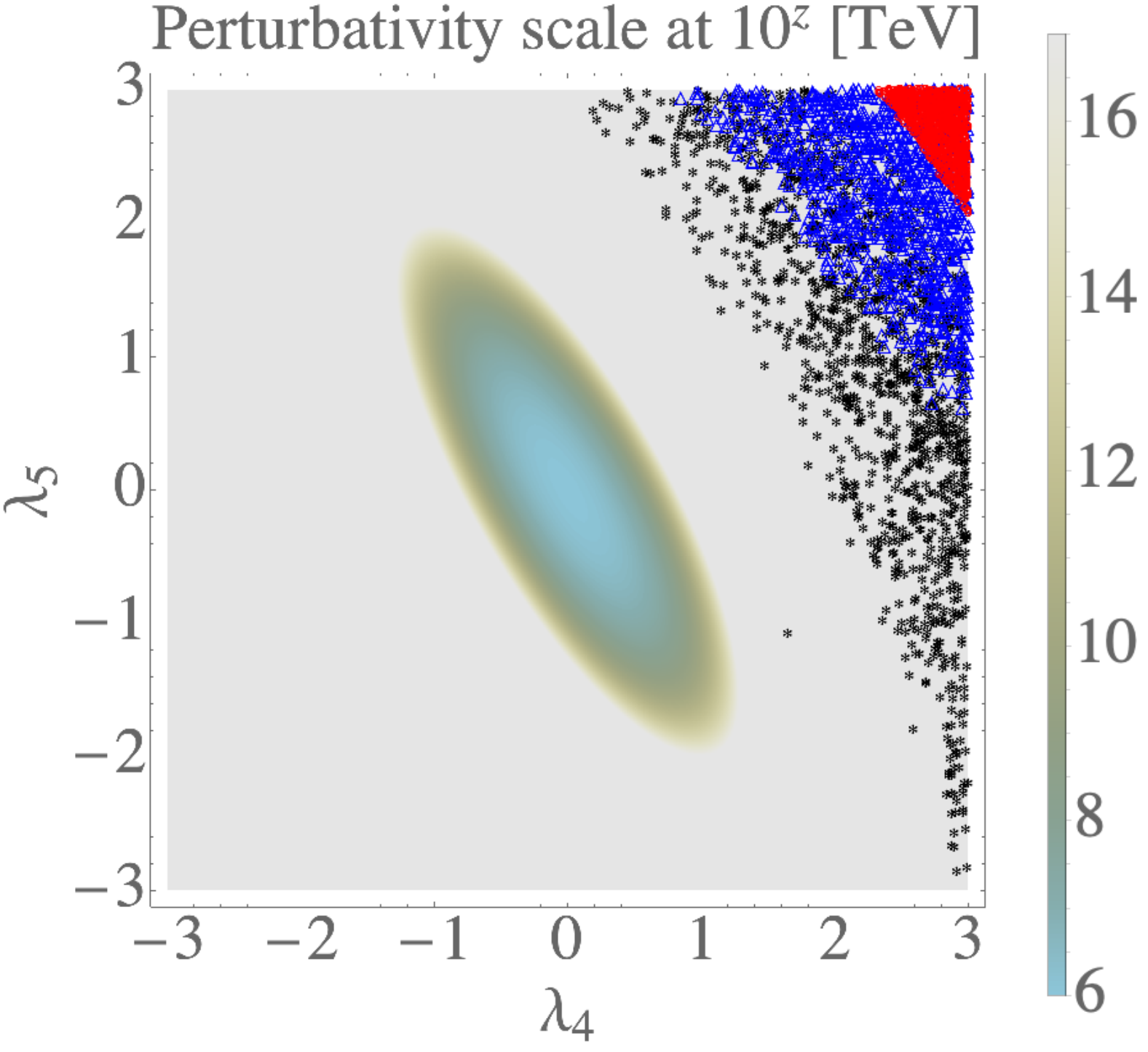}
\end{center}
\caption{Left panel: Theoretical constraints on the triplet model from tree-level vacuum stability and perturbative unitarity. Right panel: Constraints from one-loop perturbativity. In each panel, the black, blue and red points correspond to our setup one, three and four, respectively.}\label{PTAndTheCon1}
\end{figure}

To answer this question, we first map those points for each setup in section\,\ref{sebsec:nmsetup} onto the $\lambda_{4}-\lambda_{5}$ plane from considering vacuum stability, perturbative unitarity and perturbativity up to one loop. The results are shown in figure\,\ref{PTAndTheCon1}, where the left panel is from tree-level vacuum stability and perturbative unitarity, and the right one for one-loop perturbativity. The legend alongside the right panel indicates the scale to which one-loop perturbativity is satisfied. Clearly, from tree-level theoretical constraints as indicated in gray in the left panel of figure\,\ref{PTAndTheCon1}, positive $\lambda_4$ is in general preferred. This conclusion changes slightly when one-loop perturbativity is taken into account. In the latter case, requiring perturbativity up to the Planck scale, we find $\lambda_{4,5}$ with opposite signs near the origin are generically disfavored as implied by the elliptical region in the right panel.

The viable points that can lead to a SFOEWPT are then shown in black, blue, and red for setup {1}, setup {3} and setup {4}, respectively. Note that all the points fulfill tree-level constraints from perturbative unitarity and vacuum stability. Furthermore, even when one-loop perturbativity is taken into account, we find that all these points are still allowed up to the Planck scale as indicated in the right panel of figure\,\ref{PTAndTheCon1}.

\begin{figure}[!htp]
\begin{center}
\includegraphics[width=0.5\textwidth]{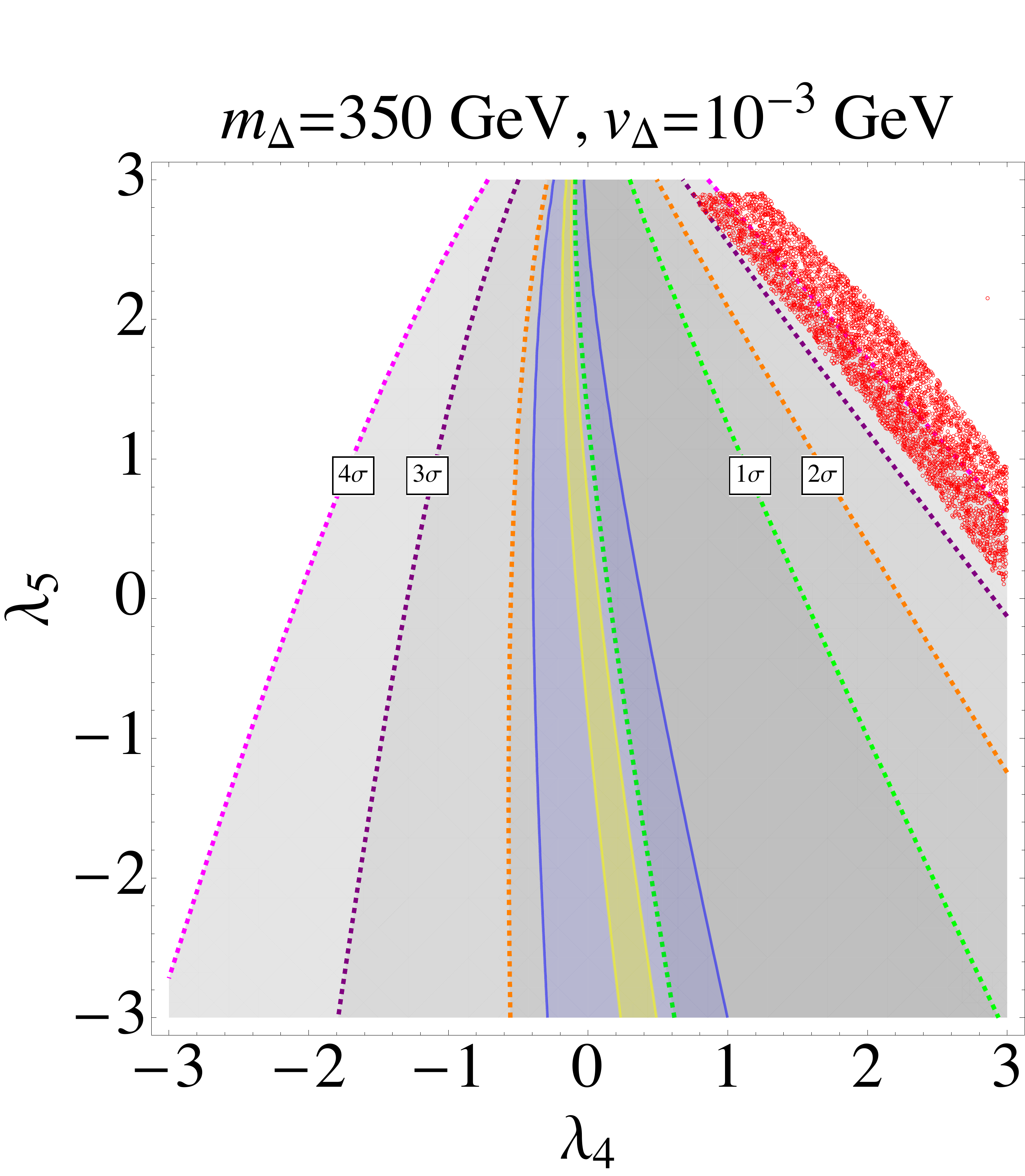}
\end{center}
\caption{Precision measurements of $\mathcal{R}_{h\gamma\gamma}$ from current and future circular colliders. The green, orange, purple and magenta boundaries represent the $1\sigma$, $2\sigma$, $3\sigma$ and $4\sigma$ regions from Ref.\,\cite{ParticleDataGroup:2020ssz}. The blue and the yellow regions are for a future 100\,TeV collider FCC-ee and FCC-ee + FCC-pp, respectively. Our benchmark points are shown in the red.}\label{PTAndTheCon2}
\end{figure}
On the other hand, from figures\,(\ref{fig_setup1Q174})-(\ref{fig_setup4Q174}), we note that to have a successful SFOEWPT, the triplet masses are generically light such that they might be within the reach of current and future colliders. For instance, when the triplet is below $\sim$1\,TeV, the same-sign dilepton (di-$W$ boson) channel would be the smoking-gun signature for discovering this model at colliders\,\cite{Du:2018eaw} for small (large) $v_\Delta$. Therefore, to answer the question we raise earlier \yong{in this section}, we make use of precision measurements of the $h\to\gamma\gamma$ decay rate defined as $\mathcal{R}_{h\gamma\gamma}\equiv\Gamma_{h\to\gamma\gamma}^{\rm NP}/\Gamma_{h\to\gamma\gamma}^{\rm SM}$, where $\Gamma_{h\to\gamma\gamma}^{\rm NP(SM)}$ is the $h\to\gamma\gamma$ decay rate with (without) the inclusion of new physics. Our benchmark scenario is obtained by fixing $m_\Delta$ at 350\,GeV and $v_\Delta\simeq10^{-3}$\,GeV, and the result is shown in figure\,\ref{PTAndTheCon2}. The shaded region in gray corresponds to $\mathcal{R}_{h\gamma\gamma}=1.11^{+0.10}_{-0.09}$ from the most recent report of PDG\,\cite{ParticleDataGroup:2020ssz}, whose $1\sigma$, $2\sigma$, $3\sigma$ and $4\sigma$ boundaries are given by the green, orange, purple and magenta dashed curves, respectively. For future circular colliders, we use the blue (yellow) region for a future 100\,TeV FCC-ee (FCC-ee + FCC-pp) collider with $\mathcal{R}_{h\gamma\gamma}=1\pm0.05\,(0.01)$\,\cite{Contino:2016spe}. The red circles in the upper right corner correspond to our benchmark points that can give a SFOEWPT within this setup.

Note that even though our benchmark points are still allowed within $3\sim4\sigma$ from the current measurement of $\mathcal{R}_{h\gamma\gamma}$ in\,\cite{ParticleDataGroup:2020ssz}, we expect the high-luminosity LHC and/or future colliders to scrutinize each of these benchmark points in this specific scenario, highlighting the powerfulness of precision measurements and the synergy of different probes.

\section{Gravitational waves from the triplet model}\label{sec:gravwave}
As discussed in section\,\ref{sec:phasetrans}, a SFOEWPT occurs when the temperature of the Universe drops below the critical temperature $T_c$. Gravitational waves (GWs) \yong{could} then be generated through collisions of vacuum bubbles, and the interaction between bubbles and the thermal plasma. The generated GWs \yong{would} then be \yong{possibly} observed by late time observatories such as {\color{black}LISA\,\cite{LISA:2017pwj}, TianQin\,\cite{TianQin:2015yph, Hu:2018yqb, TianQin:2020hid}, Taiji\,\cite{Hu:2017mde, Ruan:2018tsw}, DECIGO\,\cite{Seto:2001qf, Kudoh:2005as}, and BBO\,{\cite{Ungarelli:2005qb, Cutler:2005qq}}}. From this consideration, we discuss the synergy of different probes of the type-II seesaw model, and focus specifically on the observation of GWs in this section.\footnote{A similar discussion on the complementarity between colliders and phase transition for the singlet extension of the SM can be found in\,\cite{Alves:2018jsw}.}

The spectrum of GWs from a first-order phase transition can be obtained quite systematically. See for example, Refs.\,\cite{Caprini:2019egz}. Generically, the prediction of the GW spectrum depends on four key parameters: The bubble wall velocity $v_w$, the phase transition temperature, the latent heat $\Delta\rho$ released during the phase transition, \yong{the phase transition strength $\alpha$}, and the phase transition duration $\beta$. The definitions and their physical meaning of these parameters will become clear shortly, as will be discussed below.

Below the critical temperature, the phase transition would take place when at least one bubble is nucleated per horizon volume and per horizon time, which can be defined as as\,\cite{Affleck:1980ac,Linde:1981zj,Linde:1980tt}:
\begin{eqnarray}\label{eq:bn}
\Gamma\approx A(T_n)e^{-S_3/T_n}\simeq 1\;.
\end{eqnarray}
where $T_n$ is the nucleation temperature of the vacuum bubbles, and $S_3$ is the bounce action for an O(3) symmetric bounce solution that can be written as
\begin{eqnarray}
S_3(T)=\int 4\pi r^2d r\bigg[\frac{1}{2}\big(\frac{d \phi_b}{dr}\big)^2+V(\phi_b,T)\bigg]\;,
\end{eqnarray}
with $\phi_b = \phi,\delta$ in our case, and $V(\phi_b,T)$ the effective potential in eq.\,\eqref{potVeff}. The bubble nucleation events would be generated when one gets the bounce solution from solving the equations of motion for $\phi_b$:
\begin{eqnarray}
\frac{d^2\phi_b}{dr^2}+\frac{2}{r}\frac{d\phi_b}{dr}-\frac{\partial V(\phi_b)}{\partial \phi_b}=0\;,
\end{eqnarray}
with the boundary conditions being
\begin{eqnarray}
\lim_{r\rightarrow \infty}\phi_b =0 \;, \quad \quad {\left. {\frac{{d{\phi _b}}}{{dr}}} \right|_{r = 0}} = 0\;.
\end{eqnarray}

After nucleation, the phase transition proceeds through expansion and percolation of these vacuum bubbles. The percolation temperature $T_p$ is defined as the moment when the probability of the friction of a false vacuum is 0.7\,\cite{Guth:1981uk,Ellis:2018mja}:
\begin{eqnarray}
&&P[T_p]= e^{-I[T_p]} = 0.7\;,\nonumber\\
&&I[T]= \frac{4\pi v_w^3}{3} \int_T^{T_c} \frac{d\tilde{T} \Gamma(\tilde{T})}{H(\tilde{T})\tilde{T}^4} \left( \int_{T}^{\tilde{T}} \frac{d T^\prime}{H(T^\prime)} \right)^3 \;,
\end{eqnarray}
where $v_w$ is the bubble wall velocity.

For this study, we define the phase transition strength $\alpha$ as
\begin{eqnarray}
\alpha=\frac{\Delta\rho}{\rho_R}\;,
\end{eqnarray}
where the radiation energy density of the bath or the plasma background $\rho_R$ is given by
\begin{eqnarray}
\rho_R=\frac{\pi^2 g_\star T_\star^4}{30}\;,
\end{eqnarray}
with $g_\star\approx 100$ being the effective number of degrees of freedom, $T_\star$ the plasma temperature
that is approximately equivalent to the percolation temperature $T_\star\approx T_p$ for transitions without significant reheating\,\cite{Caprini:2015zlo}, and $\Delta\rho$ the latent heat from the phase transition. $\Delta\rho$ can be calculated from the difference of the energy density between the false and the true vacuum, i.e., $\Delta\rho=\rho(\phi_p,T_p)-\rho(v_p,T_p)$, where\footnote{ In our calculation, we use the latent heat by including the entropy injection from the phase transition (through the term of $\left.T\frac{d\, V}{d\,T}\right\vert_{T=T_p}$) as in Ref.~\cite{Kamionkowski:1993fg,Apreda:2001us,Grojean:2006bp,Huber:2007vva} and some other literatures, which coincides with the vacuum energy for the large supercooling phase transition case as commented in Ref~\cite{Caprini:2015zlo}. }
\begin{eqnarray}
\rho(\phi_p,T_p)&=& \left.-V(\phi,T)\right\vert_{T=T_p}+\left.T\frac{d\, V(\phi,T)}{d\,T}\right\vert_{T=T_p}\;,
\\
\rho(v_p,T_p)&=& -\left.V(h,T)\right\vert_{T=T_p}+\left.T\frac{d\, V(h,T)}{d\,T}\right\vert_{T=T_p}\;.
\end{eqnarray}
Here, we remind the reader that $\alpha$ and $\rho$ in this section represent the phase transition strength and the energy densities instead of
the mixing angle and the electroweak parameter discussed in section\,\ref{sec:modelsetup}. Finally, to characterizes the inverse time duration of the SFOEWPT, we define the parameter $\beta$ as
\begin{eqnarray}
\frac{\beta}{H_p}=\left.T\frac{d (S_3(T)/T)}{d T}\right\vert_{T=T_p}\; ,
\end{eqnarray}
with $H_p$ the Hubble constant at the percolation temperature $T_p$.

With above results, we are now ready to move to the discussion on the sources of GW generation from a first-order phase transition. In this work, we consider three sources for the production of GWs. The first one comes from the uncollided envelop of thin bubble walls during the bubble collision, while the collided thin bubble walls are assumed to disappear instantly after two bubbles overlap.\footnote{Recent studies of Refs.\cite{Ellis:2019oqb,Ellis:2020nnr} show that bubble collisions are usually negligible in transitions with polynomial potentials, which is true for this study.} This is the widely used envelop approximation that contributes to both numerical simulations~\cite{Kosowsky:1991ua,Kosowsky:1992rz,Kosowsky:1992vn,Kamionkowski:1993fg,Huber:2008hg} (see also~\cite{Child:2012qg}) and analytic estimations~\cite{Jinno:2016vai}.\footnote{Note that recent numerical simulations also found that the scalar oscillation stage would continue contributing to GW radiation, see Refs.~\cite{Cutting:2018tjt,Cutting:2020nla,Di:2020ivg}, and recent studies of Refs.~\cite{Lewicki:2020azd,Lewicki:2020jiv,Lewicki:2019gmv} show that the ageing envelope approximation led to inaccurate prediction for the spectrum.}
The dimensionless energy density spectrum is fitted to be~\cite{Huber:2008hg}
\begin{align}\label{eq:Omegaenv}
\Omega h^2_{\rm coll}(f)=1.67\times10^{-5}\left(\frac{100}{g_{*}}\right)^\frac13
\left(\frac{\beta}{H}\right)^{-2}\left(\frac{\kappa_\phi\alpha}{1+\alpha}\right)^2\frac{0.11v_w^3}{0.42+v_w^2}
\frac{3.8\left(f/f_{\rm coll}\right)^{2.8}}{1+2.8\left(f/f_{\rm coll}\right)^{3.8}},
\end{align}
where the first term in bracket accounts for the redshift effect, the second one reflects its scaling behavior, and the third one parameterizes the spectral shape of the GW radiation. The peak frequency $f_{\rm coll}$ involved in the spectral shape is fitted to be~\cite{Huber:2008hg}
\begin{align}\label{eq:fenv}
f_{\rm coll}=1.65\times10^{-5}\,\mathrm{Hz}\times\left(\frac{g_*}{100}\right)^\frac16\frac{T_\star}{100\mathrm{GeV}}
\frac{0.62}{1.8-0.1v_w+v_w^2} \left(\frac{\beta}{H_{\ast}} \right).
\end{align}

The other two sources for GW production during the EWPT we consider are: (1) the sound waves in the plasma~\cite{Hindmarsh:2013xza,Hindmarsh:2015qta}, and (2) the magnetohydrodynamic turbulence (MHD)~\cite{Hindmarsh:2013xza,Hindmarsh:2015qta}. For the former, taking the lifetime suppression factor obtained in\,\cite{Guo:2020grp},\footnote{The impact without including this factor has also been investigated in\,\cite{Guo:2021qcq}.} the energy density spectrum from the sound waves can be expressed as~\cite{Hindmarsh:2015qta},
\begin{align}\label{eq:sw}
\Omega h^2_{\rm sw}(f)=1.64 \times 10^{-6}&\times(H_*\tau_{sw})\left(\frac{\beta}{H}\right)^{-1}
\left(\frac{\kappa \alpha }{1+\alpha }\right)^2
\left(\frac{g_*}{100}\right)^{-\frac{1}{3}}\nonumber\\
&\times v_w(8\pi)^{1/3}
\left(\frac{f}{f_{\rm sw}}\right)^3 \left(\frac{7}{4+3 \left(f/f_{\rm sw}\right)^2}\right)^{7/2},
\end{align}
with $\tau_{sw}={\rm min}\left(\frac{1}{H_*},\frac{R_*}{\bar{U}_f}\right)$, $H_*R_*=v_w(8\pi)^{1/3}(\beta/H)^{-1}$. Here, $\bar{U}$ is the root-mean-square fluid velocity that can be approximated as\,\cite{Hindmarsh:2017gnf,Caprini:2019egz,Ellis:2019oqb}
\begin{equation}
\bar{U}_f^2\approx\frac{3}{4}\frac{\kappa_\nu\alpha}{1+\alpha}\;,
\end{equation}
and again, $\alpha$ here is the phase transition strength. The term $H_*\tau_{\rm sw}$ in eq.\,\eqref{eq:sw} accounts for the suppression of the GW amplitude for sound waves if the sound wave source could not last longer than one Hubble time, and $H_{\ast}$ is the Hubble parameter at the temperature $T_{\ast}$. Practically, $T_{\ast}$ is very close to $T_p$, and for this reason, we replace $T_{\ast}$ by $T_p$ in the our calculations.
$\kappa_v$ is the fraction of the released energy into the kinetic energy of the plasma, which can
be calculated given $v_w$ and $\alpha$~\cite{Espinosa:2010hh}.
Finally $f_{\text{sw}}$ is the peak frequency of above energy density spectrum:
 \begin{equation}
f_{\textrm{sw}}=1.9\times10^{-5}\frac{1}{v_{w}}\left(\frac{\beta}{H_{\ast}} \right) \left( \frac{T_{\ast}}{100\textrm{GeV}} \right) \left( \frac{g_{\ast}}{100}\right)^{1/6} \textrm{Hz} .
\end{equation}

On the other hand, for the latter source of GW production, it arises from the fact that a small fraction of the energy would flow into the MHD. Its contribution to the energy density spectrum can be expressed as~\cite{Caprini:2009yp,Binetruy:2012ze}
\eqal{
\Omega_{\textrm{turb}}h^{2}=&\,3.35\times10^{-4}\left( \frac{\beta}{H_{\ast}}\right)^{-1} \left(\frac{\kappa_{\text{turb}}
\alpha}{1+\alpha} \right)^{3/2} \left( \frac{100}{g_{\ast}}\right)^{1/3}\nonumber\\
&\,\times v_{w} \cdot \frac{(f/f_{\textrm{turb}})^{3}}{[1+(f/f_{\textrm{turb}})]^{11/3}(1+8\pi f/h_{\ast})} \;, \quad
\label{eq:mhd}
}
where the factor $\kappa_{\text{turb}}$ is the fraction of energy transferred to the MHD turbulence and can be roughly estimated as $\kappa_{\text{turb}} \approx \epsilon \kappa_{v}$ with 
$\epsilon\approx$ 5 $\sim$ 10\%~\cite{Hindmarsh:2015qta}. In this work, we take $\epsilon\approx 0.1$ for the following discussion.
Similar to $f_{\text{sw}}$, $f_{\text{turb}}$ is the peak frequency for the spectrum from the MHD:
\begin{equation}
f_{\textrm{turb}}=2.7\times10^{-5}\frac{1}{v_{w}}\left(\frac{\beta}{H_{\ast}} \right) \left( \frac{T_{\ast}}{100\textrm{GeV}} \right) \left( \frac{g_{\ast}}{100}\right)^{1/6} \textrm{Hz} .
\end{equation}

\begin{table}[!tbp]
\begin{center}
\begin{tabular}{c| c c c c c c }
\hline
&~~~~&$T_p$(GeV)~~&~~$\alpha[T_p]$~~&~~$\beta/H[T_p]$ \\
\hline
setup {1}&${\rm BM}_1$ &96.701&0.048&657.743\\
setup {3}&${\rm BM}_2$ &99.195&0.046&1026.894\\
setup {4}&${\rm BM}_3$ &136.708&0.015&2712.428\\
\hline
\end{tabular}\caption{Three benchmark points for the illustration of GW production from a SFOEWPT for the setup 1, 3, and 4. Setup 2 is missing in this table due to the decoupling.}\label{tab:data}
\end{center}
\end{table}

\begin{figure}[t]
\centering{
  \begin{adjustbox}{max width = \textwidth}
\begin{tabular}{cc}
\includegraphics[width=0.51\textwidth]{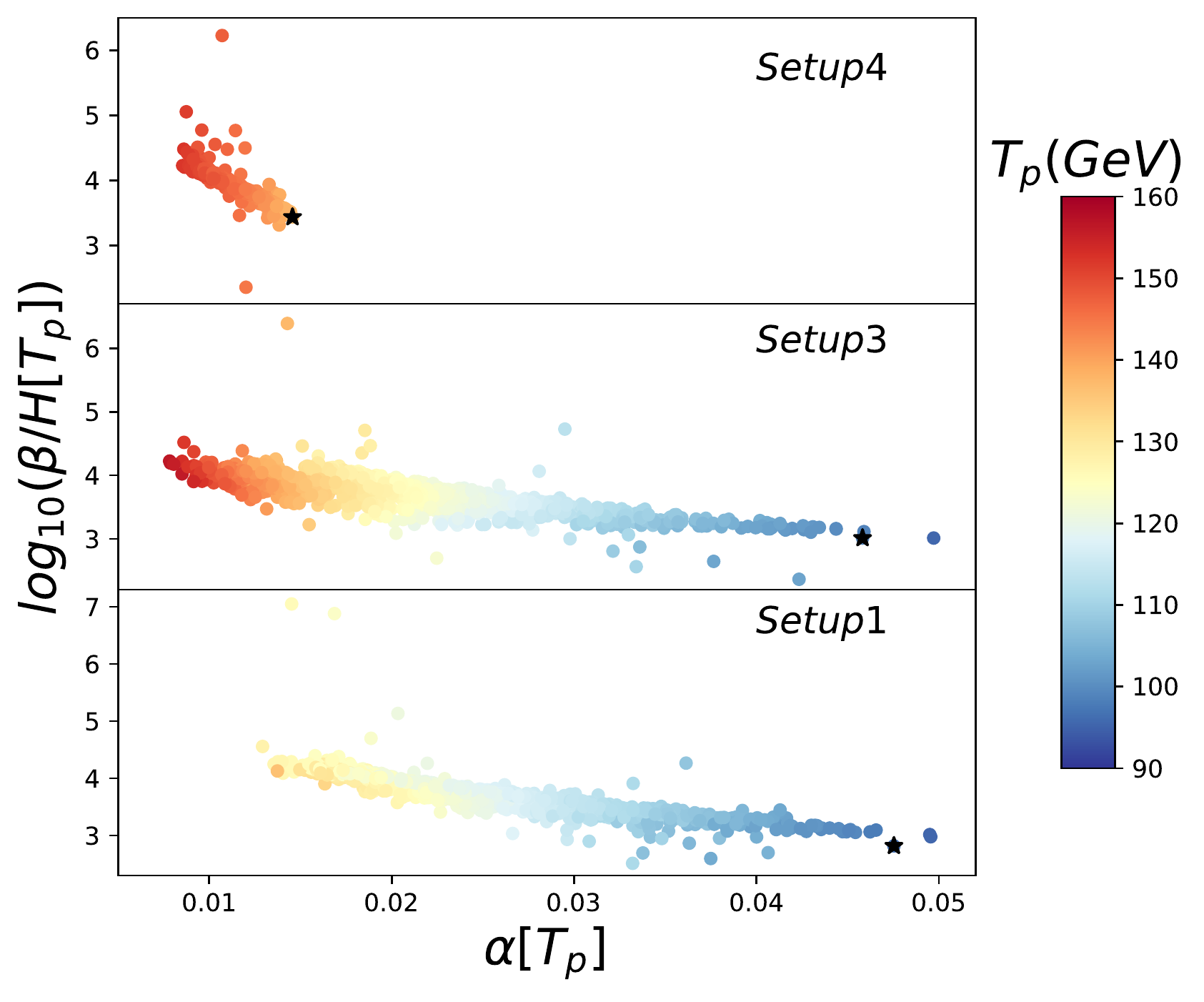}&\includegraphics[width=0.43\textwidth]{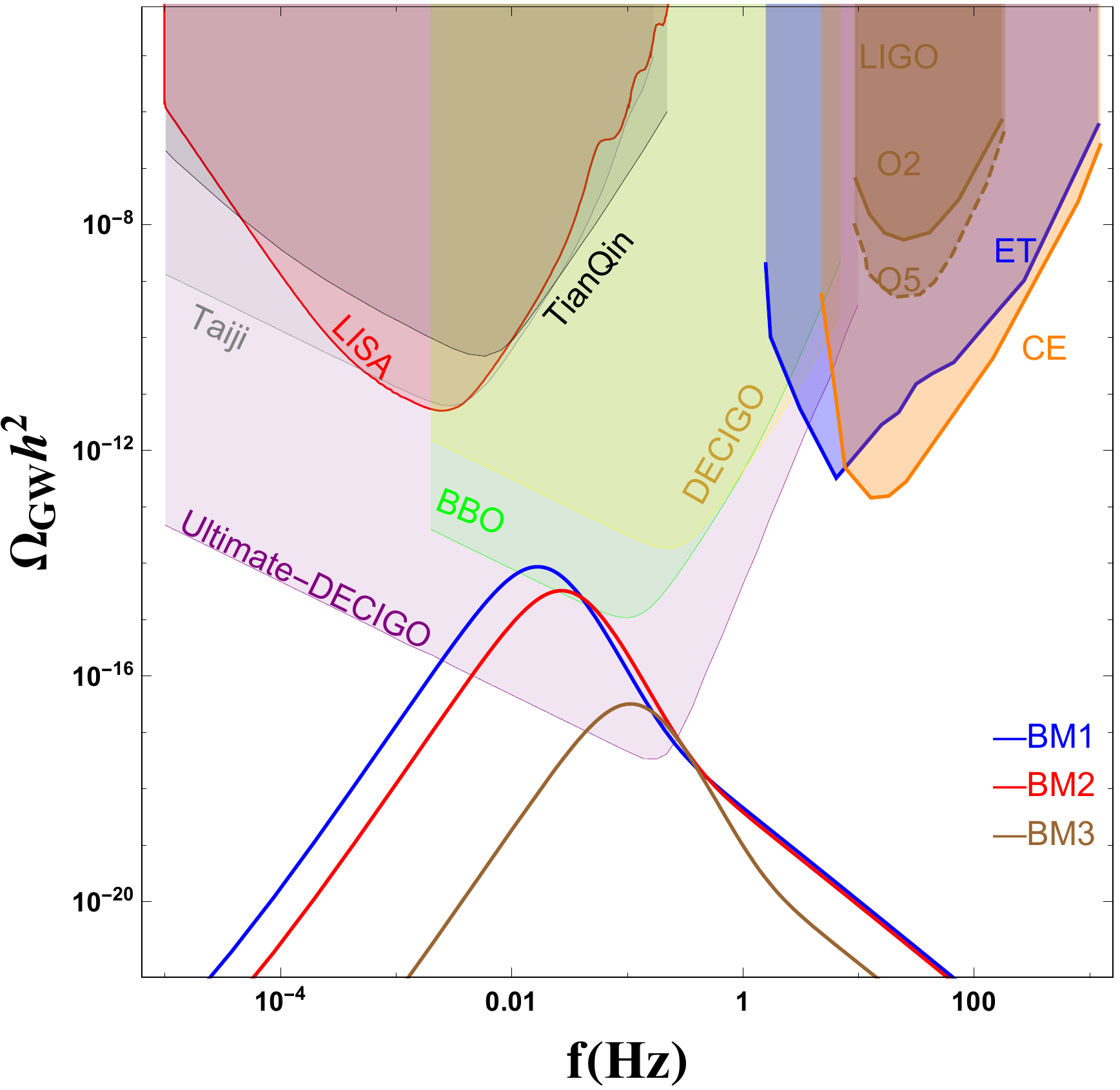}
\end{tabular}
\end{adjustbox}
}
\caption{Left: $\beta/H[T_p]$ and $\alpha$ for the four benchmark setups in section\,\ref{sebsec:nmsetup}. Note that setup 2 is missing due to the decoupling of the triplet. The legend along with the figure is used to indicate the percolation temperature $T_p$, and the black stars denote three benchmark points in table\,\ref{tab:data}. Right: The representative GW signal spectrum for the three benchmark points in table\,\ref{tab:data}. See the main text for details.}\label{fig_GW}
\end{figure}

The predicted GW spectrum can then be readily calculated from the three sources discussed above, leading to
\eqal{\Omega_{\rm GW}h^2  = \Omega h^2_{\rm coll}(f) + \Omega h^2_{\rm sw}(f) + \Omega_{\textrm{turb}}h^{2}\;.}
This predicted spectrum could then be tested at various GW observatories mentioned above, thus it could also be used for discovering/testing specific UV models like the type-II seesaw model considered in this work. To that end, we choose three benchmark points for the four setups discussed in section\,\ref{sebsec:nmsetup} and comment on the fact that no benchmark points are selected for the second setup due to the decoupling discussed earlier. The selected benchmark points for the rest three setups are then summarized in table\,\ref{tab:data}, whose effective potential have been presented in figure\,\ref{fig:veff} and their corresponding results for GWs are presented in figure\,\ref{fig_GW}.

In the left panel of figure\,\ref{fig_GW}, we show the results for $\beta/H[T_p]$ and $\alpha[T_p]$ for varying percolation temperatures. Note that, as self-explained in eqs.\,\eqref{eq:Omegaenv}, \eqref{eq:sw}, and \eqref{eq:mhd}, the magnitude of GWs is inversely proportional to $\beta/H_\star$ and directly proportional to $\alpha$ for fixed $v_w$ and $T_\star$. As a result, one naturally expects that a larger value of $\alpha$ and/or a smaller value of $\beta/H_\star$ would lead to an increase in the magnitude of the GWs observed. This is as expected since a larger value of $\alpha$ would suggest more energy transition from the plasma to the form of GWs. Similarly, a smaller $\beta/H_\star$ would imply a longer period for the strong first-order phase transition, thus also enhancing the magnitude of the spectrum. This is also confirmed numerically as shown in the right panel of figure\,\ref{fig_GW}.

The predicted spectra for the three benchmark points in table\,\ref{tab:data} are presented in blue, red and orange in the right panel of figure\,\ref{fig_GW}, respectively. These three benchmarks are chosen with relatively large $\alpha$ and small $\beta/H_\star$ from the left panel of figure\,\ref{fig_GW} to enhance the magnitude of the generated GWs. See the corresponding black stars in the left panel for these three benchmark points we choose. As a result, we find the generated GW waves from the SFOEWPT all have a peak frequency within the $0.01\sim0.1$Hz range, with the peak yields of the GWs around $10^{-14}$, $10^{-15.5}$ and $10^{-16.5}$ for our BM1 (blue), BM2 (red), and BM3 (brown), respectively. From the right panel of figure\,\ref{fig_GW}, we comment on that while the peak yields of the GWs for our BM2 and BM3 are small, they would be covered by Ultimate-DECIGO in the future. In particular, we see that BBO \yong{would be able to explore} the edge of the BM1 scenario, and the Ultimate-DECIGO would have the chance to further examine both the BM1 and the BM2 cases.

\section{Conclusions}\label{sec:conclu}
Neutrino masses and the baryon asymmetry of the Universe both indicate new physics beyond the SM. In this work, we focus on the type-II seesaw model that acts as a possible candidate for answering these two questions simultaneously. Specifically, the type-II seesaw model can be obtained by extending the SM Higgs sector with a complex triplet that transforms as $(1,3,2)$ under the SM gauge group. Due to the quantum numbers of the triplet, new interactions are introduced between the SM Higgs doublet and the complex triplet, such that the SM Higgs potential could be modified in a way such that a SFOEWPT is possible.

We study the phase transition within the triplet model in detail in this work and obtain viable regions of the model parameter space for a SFOEWPT that is responsible for explaining the observed baryon asymmetry. Our results are shown in figures\,\ref{fig:sfoewpt}-\ref{fig_setup4Q174} for the four setups discussed in section\,\ref{sec:phasetrans}. We find that when the triplet is heavy above $\sim550$\,GeV, effects on the Higgs potential from the triplet would decouple such that a SFOEWPT would become absent in this model. Furthermore, we conclude from our study that a SFOEWPT generically prefers positive values for the Higgs portal couplings $\lambda_{4,5}$, which in turn help stabilize the Higgs potential up to the Planck scale up to one-loop level\,\cite{Du:2022vso}. We point out that the Higgs di-photon decay rate is also sensitive to $\lambda_{4,5}$\,\cite{Du:2018eaw}, such that a precision measurement on the rate could shed some light on the phase transition. This highlights the synergy of different probes in searching for new physics.

On the other hand, gravitational waves can be also generated during the SFOEWPT from bubble collisions and its interaction with thermal plasma. This has been investigated in section\,\ref{sec:gravwave} in the complex triplet model, and the results are presented in our figure\,\ref{fig_GW}. For the four setups we consider that cover the model parameter space up to 4\,TeV, we obtain the \yong{phase transition strength} $\alpha$ and the phase transition duration $\beta$ for various percolation temperatures. Based on that, we then choose three optimistic benchmark points to calculate the gravitational wave yields and compare them with different observatories now and in the future. We find the peak frequency of the gravitational waves could be within the 0.01$\sim$0.1 Hz range, with a peak yield of gravitational waves at the edge of BBO and could be further examined in the future by Ultimate-DECIGO.

Last but not least, we comment that, for a successful first-order phase transition and a relatively large yield of gravitational waves, we observe that the triplet Higgs particles are preferred to be light below the TeV scale. With the triplet particles being light at such a scale, the triplet vev would need to be large above $\sim10^{-4}$\,GeV to avoid very stringent constraints from current collider searches\,\cite{ATLAS:2017xqs,Du:2018eaw}. This in turn would result in tiny neutrino Yukawa couplings due to the tininess of neutrino masses. As a consequence, one would thus expect the triplet model not to manifest itself in foreseen neutrino oscillation experiments due to the neutrino Yukawa suppression. However, collider searches would help with the same-sign di-$W$ boson final state being the smoking-gun signature.

\acknowledgments{
This work was supported in part by National Key Research and Development Program of China Grant Nos. 2020YFC2201501, 2021YFC2203004. Ligong Bian was supported by the National Natural Science Foundation of China under the grants Nos.12075041, 12047564, and the Fundamental Research Funds for the Central Universities of China (No. 2021CDJQY-011, No. 2020CDJQY-Z003, and No. 2021CDJZYJH-003), and Chongqing Natural Science Foundation (Grants No.cstc2020jcyj-msxmX0814).
Yong Du was supported in part by the National Science Foundation of China (NSFC) under Grants No. 12022514, No. 11875003 and No. 12047503, and CAS Project for Young Scientists in Basic Research YSBR-006, and the Key Research Program of the CAS Grant No. XDPB15.   }

\bibliographystyle{JHEP}
\bibliography{ref}

\end{document}